\documentclass[%
 reprint,
superscriptaddress,
nofootinbib,
 amsmath,amssymb,
 aps,
prd,
floatfix,
]{revtex4-2}

\usepackage{graphicx}
\usepackage{dcolumn}
\usepackage{bm}
\usepackage[table,svgnames,dvipsnames]{xcolor}
\usepackage[mathlines]{lineno}
\usepackage[normalem]{ulem}
\usepackage{aas_macros}
\usepackage{subfigure}
\usepackage{graphicx}
\usepackage{graphics}
\usepackage{dcolumn}
\usepackage{bm}
\usepackage{cases}
\usepackage{booktabs}
\usepackage{comment}
\usepackage{multirow}
\usepackage{makecell}
\usepackage{siunitx}
\usepackage{tabularx}
\usepackage{xspace}
\usepackage{soul} 
\graphicspath{{figs/}}
\usepackage{fontawesome}
\usepackage{hyperref}

\usepackage{orcidlink} 
\usepackage{hanging} 

\def\be{\begin{equation}}
\def\ee{\end{equation}}

\def\ba#1\ea{\begin{align*}#1\end{align*}}

\renewcommand{\emph}[1]{\textit{#1}}
\definecolor{RoyalBlue}{rgb}{0.25,.41,.88}
\definecolor{WildStrawberry}{HTML}{EE2967}
\definecolor{RedWine}{rgb}{0.743,0,0}
\definecolor{bittersweet}{rgb}{1.0, 0.44, 0.37}
\definecolor{burntorange}{rgb}{0.8, 0.33, 0.0}
\definecolor{midnightgreen}{rgb}{0.0, 0.29, 0.33}
\definecolor{otherblue}{rgb}{0.20, 0.73, 0.92}
\definecolor{purple}{rgb}{0.647,0.439,0.882}
\definecolor{UltraViolet}{HTML}{6433FF}

\newcommand{\DVrd}{D_\mathrm{V}/r_\mathrm{d}}
\newcommand{\DMrd}{D_\mathrm{M}/r_\mathrm{d}}
\newcommand{\DHrd}{D_\mathrm{H}/r_\mathrm{d}}
\newcommand{\DM}{D_\mathrm{M}}
\renewcommand{\DH}{D_\mathrm{H}}

\usepackage[nameinlink,noabbrev]{cleveref}
\crefname{equation}{Eq.}{Eqs.}
\crefname{section}{Section}{Sections}
\crefname{figure}{Figure}{Figures}
\crefname{table}{Table}{Tables}
\crefname{appendix}{Appendix}{Appendices}
\Crefname{figure}{Figure}{Figures}
\Crefname{equation}{Equation}{Equations}
\Crefname{section}{Section}{Sections}
\Crefname{table}{Table}{Tables}

\newcommand{\mksym}[1]{\ifmmode {\rm #1}\else #1\fi}

\newcommand{\Om}{\Omega_\mathrm{m}}

\newcommand{\ob}{\omega_\mathrm{b}}
\newcommand{\ocdm}{\omega_\mathrm{cdm}}

\newcommand{\Ode}{\Omega_\mathrm{DE}}

\newcommand{\lcdm}{$\Lambda$CDM}

\newcommand{\DV}{D_\mathrm{V}}

\newcommand{\rd}{r_\mathrm{d}}
\newcommand{\zd}{z_\mathrm{d}}

\newcommand{\fde}{\ensuremath{f_\mathrm{DE}}}
\newcommand{\chisq}{\ensuremath{\chi^2}}
\newcommand{\dchisq}{\ensuremath{\Delta\chi^2}}
\newcommand{\dchisqMAP}{\Delta\chi^2_\mathrm{MAP}}

\def\tpdf#1{\texorpdfstring{#1}{Lg}}

\begin{document}

\title{
Extended Dark Energy analysis using DESI DR2 BAO measurements
}


\author{K.~Lodha\orcidlink{0009-0004-2558-5655}}
\email{kushallodha@kasi.re.kr}
\affiliation{Korea Astronomy and Space Science Institute, 776, Daedeokdae-ro, Yuseong-gu, Daejeon 34055, Republic of Korea}
\affiliation{University of Science and Technology, 217 Gajeong-ro, Yuseong-gu, Daejeon 34113, Republic of Korea}

\author{R.~Calderon\orcidlink{0000-0002-8215-7292}}
\affiliation{CEICO, Institute of Physics of the Czech Academy of Sciences, Na Slovance 1999/2, 182 21, Prague, Czech Republic.}

\author{W.~L.~Matthewson\orcidlink{0000-0001-6957-772X}}
\affiliation{Korea Astronomy and Space Science Institute, 776, Daedeokdae-ro, Yuseong-gu, Daejeon 34055, Republic of Korea}

\author{A.~Shafieloo\orcidlink{0000-0001-6815-0337}}
\affiliation{Korea Astronomy and Space Science Institute, 776, Daedeokdae-ro, Yuseong-gu, Daejeon 34055, Republic of Korea}
\affiliation{University of Science and Technology, 217 Gajeong-ro, Yuseong-gu, Daejeon 34113, Republic of Korea}

\author{M.~Ishak\orcidlink{0000-0002-6024-466X}}
\affiliation{Department of Physics, The University of Texas at Dallas, 800 W. Campbell Rd., Richardson, TX 75080, USA}

\author{J.~Pan\orcidlink{0000-0001-9685-5756}}
\affiliation{University of Michigan, 500 S. State Street, Ann Arbor, MI 48109, USA}

\author{C.~Garcia-Quintero\orcidlink{0000-0003-1481-4294}}
\affiliation{Center for Astrophysics $|$ Harvard \& Smithsonian, 60 Garden Street, Cambridge, MA 02138, USA}
\affiliation{NASA Einstein Fellow}

\author{D.~Huterer\orcidlink{0000-0001-6558-0112}}
\affiliation{University of Michigan, 500 S. State Street, Ann Arbor, MI 48109, USA}
\affiliation{Department of Physics, University of Michigan, 450 Church Street, Ann Arbor, MI 48109, USA}

\author{G.~Valogiannis\orcidlink{0000-0003-0805-1470}}
\affiliation{Department of Astronomy and Astrophysics, University of Chicago, 5640 South Ellis Avenue, Chicago, IL 60637, USA}
\affiliation{Fermi National Accelerator Laboratory, PO Box 500, Batavia, IL 60510, USA}

\author{L.~A.~Ure\~na-L\'opez\orcidlink{0000-0001-9752-2830}}
\affiliation{Departamento de F\'{\i}sica, DCI-Campus Le\'{o}n, Universidad de Guanajuato, Loma del Bosque 103, Le\'{o}n, Guanajuato C.~P.~37150, M\'{e}xico}

% ...

\collaboration{DESI Collaboration}

\begin{abstract}

\newpage
We conduct an extended analysis of dark energy constraints, in support of the findings of the DESI DR2 cosmology key paper, including DESI data, Planck CMB observations, and three different supernova compilations. Using a broad range of parametric and non-parametric methods, we explore the dark energy phenomenology and find consistent trends across all approaches, in good agreement with the $w_0w_a$CDM key paper results. Even with the additional flexibility introduced by non-parametric approaches, such as binning and Gaussian Processes, we find that extending $\Lambda$CDM to include a two parameter $w(z)$ is sufficient to capture the trends present in the data. Finally, we examine three dark energy classes with distinct dynamics, including quintessence scenarios satisfying $w \geq -1$, to explore what underlying physics can explain such deviations. The current data indicate a clear preference for models that feature a phantom crossing; although alternatives lacking this feature are disfavored, they cannot yet be ruled out. Our analysis confirms that the evidence for dynamical dark energy, particularly at low redshift ($z \lesssim 0.3$), is robust and stable under different modeling choices.
\end{abstract}

\maketitle

\tableofcontents

\section{Introduction}

\label{sec:intro}

The $\Lambda$ Cold Dark Matter ($\Lambda$CDM) model has withstood the test of time as the standard framework of modern cosmology, and it provides a robust foundation for understanding the Universe. It describes a spatially flat universe that is homogeneous and isotropic on large scales, governed by Einstein’s general relativity~\cite{Einstein:1917ce}. The model incorporates two key components: about 70\% is in dark energy that is described by the vacuum-energy contribution (corresponding to the cosmological constant $\Lambda$ in the equations), while another 30\% is in pressureless matter that is made up of a combination of cold dark matter (CDM) and baryons.  
Despite its elegant simplicity, $\Lambda$CDM has successfully explained a broad range of cosmological observations~\cite{SupernovaSearchTeam:1998fmf,SupernovaCosmologyProject:1998vns,Percival2002:astro-ph/0206256,2005Eisenstein,20052dFBAO,Planck-2018-cosmology,Alam_2021,Zhao_2022,DES:2017myr,DES:2017qwj,eBOSS:2020yzd,Heymans:2020gsg,DES:2021wwk}. On the whole, measurements made over the past several decades have largely confirmed this paradigm and, in particular, cemented dark energy \cite{Efstathiou:1999,Frieman:2008sn,Weinberg:2013agg} 
as the essential component of concordance model to explain the observed accelerated expansion of the Universe \cite{SupernovaSearchTeam:1998fmf,SupernovaCosmologyProject:1998vns}.

While the cosmological constant has been a cornerstone of the standard model of cosmology, various dark energy models with an evolving equation of state have been proposed as alternatives~\cite{Ostriker:1995su, RevModPhys.61.1,Ratra:1987rm, Peebles:1987ek,Sahni:1999gb,Peebles:2002gy,Copeland:2006wr,Bull:2015stt,Perivolaropoulos:2021jda}.
Specifically, we are motivated to study these time-evolving alternatives by the recent cosmological results from the Dark Energy Spectroscopic Instrument (DESI) \cite{Snowmass2013.Levi,DESI2016b.Instr}. DESI is able to measure multiple spectra simultaneously by means of its 5,000 fibers \cite{FiberSystem.Poppett.2024} and a robotic plane assembly \cite{FocalPlane.Silber.2023} across the field of view given its $3.2^\circ$ diameter prime focus corrector \cite{Corrector.Miller.2023}. This is complemented by a high-performance spectroscopic data processing pipeline \cite{Spectro.Pipeline.Guy.2023} and a streamlined operations plan \cite{SurveyOps.Schlafly.2023}. DESI is designed to help better understand the nature of dark energy \cite{DESI2016a.Science} and its successful survey validation \cite{DESI2022.KP1.Instr} based on early data \cite{DESI2023b.KP1.EDR} showed that it meets the expected requirements of a Stage-IV survey. In particular, its Data Release 1 (DR1 \cite{DESI2024.I.DR1}) has already provided new insights into the behavior of dark energy. DESI DR1 measured the baryon acoustic oscillations (BAO) signature in the clustering of galaxies and quasars \cite{DESI2024.III.KP4}, as well as the Lyman-$\alpha$ forest \cite{DESI2024.IV.KP6}.  The combined constraints from DESI DR1 BAO and external data \cite{DESI2024.VI.KP7A}, followed up with a similar analysis that combines the BAO with the full clustering information from DESI galaxies and other tracers \cite{DESI2024.VII.KP7B,DESI2024.V.KP5}, as well as the supporting DESI DR1 papers that considered alternative descriptions of the dark energy sector \cite{DESI:2024aqx,DESI:2024kob}, all showed tantalizing hints of the departures from the cosmological constant dark energy model. 
Cosmological hints in the dark energy sector are currently a source of debate, and it is of high priority to explore them with more data. In this work, we make use of the BAO measurements from the second data release (DR2 \cite{DESI.DR2.DR2,DESI.DR2.BAO.lya,DESI.DR2.BAO.cosmo}) from DESI to explore the possibility of an evolving, dynamical dark energy, and evaluate whether existing observations support such a paradigm shift.
This paper is part of a set of supporting papers that aim to extend the cosmological analysis presented in \cite{DESI.DR2.BAO.cosmo} (see \cite{Y3.cpe-s2.Elbers.2025} for the supporting paper focusing on neutrino constraints).

An essential ingredient, in a study confronting dark-energy models with data, is the physical description of dark energy (DE). In the standard concordance model, \lcdm, it is described by its contribution to the stress-energy tensor, $\Lambda$ or, equivalently, by its energy density relative to critical, $\Ode$. A dynamical dark energy is enabled by allowing the equation of state, $w=P/(\rho c^2)$, to differ from its \lcdm\ value of $-1$. 
There are many possible ways to achieve this, a large number of which have been introduced and tested in the literature \citep{ Huterer:2000mj,CHEVALLIER_2001,Linder2003,Huterer_2003,Shafieloo:2005nd,dePutter2008,2009JCAP...12..025C,Bogdanos:2009ib,Holsclaw:2010,Holsclaw_2011,Zhao:2012aw,Nesseris_2012,LHuillier:2016mtc,Calderon:2022cfj}. We can classify them as parametric and non-parametric approaches. Parametric approaches rely on predefined functional forms for quantities like $w(z)$ (where $z$ is redshift), while non-parametric methods seek to reconstruct these quantities directly from data without assuming predefined functional forms or specific cosmological models. Both methods have advantages and disadvantages. On the one hand, parametric models are mathematically simple and easy to interpret, but may lead to biased inferences if the assumed parametric form deviates substantially from reality. On the other hand, non-parametric methods offer greater flexibility and are less subject to model-dependent biases. However, these are harder to implement and require careful validation with simulations.
For this reason, we perform initial tests using simulated (mock) datasets. While there is no substitute for comprehensive validation, these tests check the methodology's implementation and mitigate potential biases that could affect the results. 
We remind readers that all the analyses in this paper rely on the assumption that the data used are reliable and free from unknown systematics.

The paper is structured as follows. In~\cref{sec:data}, we introduce the datasets and general methodology used in the analyses, followed by~\cref{sec:Overview}, where we summarize the current status of the DESI results from the $w_0w_a$ parametrization~\cite{DESI.DR2.BAO.cosmo}. Various alternative dark energy parametrizations are explored in~\cref{sec:params}, before the implementation of non-parametric methods in~\cref{sec:non-params}. ~\cref{sec:implications} provides an interpretation of the possible physical mechanisms behind deviations from $\Lambda$CDM. Finally, in~\cref{sec:conclusions}, we present our conclusions.

\section{Datasets and Methodology}
\label{sec:data}
In this section, we provide a brief cosmological background on distance measurements relevant to DESI, with an emphasis on dark energy. 
We start by introducing the relevant cosmological functions, before proceeding to describe the datasets used and the statistical tools employed in our analysis. 

As shown in \cite{DESI2024.VI.KP7A}, the evidence for spatial curvature in the Universe is not significant. Therefore, we assume a flat universe for all the results presented in our analyses.  The time-dependence of the dark energy density is enabled via the equation of state $w(z)$; the expansion rate reads

\begin{equation}
    \begin{aligned}
    \frac{H(z)}{H_0} = \Big[&\Omega_{\rm bc}(1+z)^3 + \Omega_{\gamma} (1+z)^4 + \Omega_\nu \frac{\rho_\nu(z)}{\rho_{\nu,0}} \\
    &+ 
    \Ode \frac{\rho_{\mathrm{DE}}(z)}{\rho_{\mathrm{DE},0}} \Big]^{1/2}\;,
     \label{eq:friedmann}
\end{aligned}
\end{equation}
where $\Omega_{\rm bc} = \Omega_{\rm c} + \Omega_{\rm b}$, $H_0$ is the Hubble parameter today,  and  $\Omega_{\rm b}$,  $\Omega_{\rm c}$, $\Omega_\gamma$, $\Omega_\nu$ and $\Ode$ are the present-time energy density parameters in baryons, cold dark matter, radiation, massive neutrinos and dark energy, respectively. 
The neutrino species contribute to the matter content of the Universe at the present day, since they behave as non-relativistic matter once they have passed through a transition redshift during the matter domination era (e.g., transitioning around a redshift $\sim 100$ for a neutrino mass eigenstate with a mass of 0.06 eV)~\cite{PDG:2022,Lesgourgues:2006}. This detail will be important when defining our cosmic microwave background (CMB) compression scheme, since relativistic neutrinos do not contribute significantly to matter density at the time of recombination.   We define $\Om = \Omega_{\rm bc} + \Omega_{\nu}$ to denote the matter content that scales $(1+z)^3$ when neutrinos are non-relativistic.

For a dark energy component with an equation-of-state parameter $w(z) = P_\mathrm{DE}(z)/(\rho_\mathrm{DE}(z) c^2)$, the  energy density $\rho_{\rm DE}$ normalized to its present value evolves as 
\begin{equation}
    f_{\rm DE}(z) \equiv \frac{\rho_\mathrm{DE}(z)}{\rho_\mathrm{DE,0}} = \exp\left[3\int_0^z [1+w(z^\prime)] \frac{dz^\prime}{1+z^\prime}\right] ~.
  \label{eq:DEevolution}
\end{equation}
For a constant value of $w(z)$, the dark energy density becomes proportional to $(1+z)^{3(1+w)}$, while for a model based a cosmological constant ($w=-1$), the right-hand side of \cref{eq:DEevolution} is unity. 
The conventional $w_0w_a$ parametrization for time-varying $w(z)$ is~\cite{CHEVALLIER_2001,Linder2003}
\begin{equation}
    w(z)=w_0+w_a\,\frac{z}{1+z}\,
    \label{eq:w0wa}
\end{equation} 
with energy density following the expression:
\begin{equation}
    \fde(z) = (1+z)^{3(1+w_0+w_a)}e^{-3w_a \frac{z}{1+z}}\,.
    \label{eq:w0wa_fde}
\end{equation} 

\clearpage

\begin{table}
    \centering
    \caption{
    Parameters and priors used in the analysis. 
    In addition to the flat priors on $w_0$ and $w_a$ listed in the table, we also impose the requirement $w_0+w_a<0$ in order to enforce a period of high-redshift matter domination.
    }
    \begin{ruledtabular}
    \begin{tabular}{llll}
    parametrization & parameter & default & prior\\  
    \hline 
    \textbf{Baseline} & $\ocdm$ &---& $\mathcal{U}[0.001, 0.99]$ \\   
    & $\ob$ &---& $\mathcal{U}[0.005, 0.1]$ \\
    & $100 \theta_{\mathrm{MC}}$ &---& $\mathcal{U}[0.5, 10]$ \\
    & $\ln(10^{10} A_{s})$ &---& $\mathcal{U}[1.61, 3.91]$ \\
    & $n_{s}$ &---& $\mathcal{U}[0.8, 1.2]$ \\
    & $\tau$ &---& $\mathcal{U}[0.01, 0.8]$ \\
    in absence of $\theta_{\rm MC} $ & $H_{0} \;$ &---& $\mathcal{U}[20, 100]$  \\
    \hline 
    \textbf{Alt. Parametrization} & $w_0$ & $-1$ & $\mathcal{U}[-3, 1]$ \\
    & $w_{a}$ & $0$ & $\mathcal{U}[-3, 2]$ \\
    \hline
    \textbf{Crossing} & $C_0$ & $1$ & $\mathcal{N}[1, 1^2]$ \\
    & $C_i$ & $0$ & $\mathcal{N}[0, 1^2]$ \\
    \hline
    \textbf{Binning} & $w_i$ & $-1$ & $\mathcal{U}[-3, 1]$ \\
    & $\fde{}_{,i}$ & $1$ & $\mathcal{U}[-5, 5]$ \\
    \hline
    \textbf{Gaussian Processes} & $\ell_f$ & --- & \cref{eq:ellf_prior} \\
    & $w(z_l)$ & $-1$ & $\mathcal{N}[-1, 1^2]$ \\
    \hline
    \textbf{Dark Energy Classes} &  &  & \\
    Calib. Thawing & $w_0$ & --- & $\mathcal{U}[-3, 1]$ \\
    Algebraic Thawing & $w_0$ & --- & $\mathcal{U}[-1, 1]$ \\
     & $p$ & --- & $\mathcal{U}[0,30 ]$ \\
    Emergent & $\Delta$ & --- & $\mathcal{U}[-3, 10]$ \\
    Mirage & $w_0$ & --- & $\mathcal{U}[-3, 1]$ \\
    \end{tabular}
    \end{ruledtabular}
    \label{tab:priors}
\end{table}

BAO measure the comoving distance at the effective redshift of a given galaxy sample, in units of the sound horizon ($r_{\rm s}$) at the drag epoch, labeled as $r_{\rm d} \equiv r_{\rm s}(z_{\rm d})$. The drag epoch corresponds to the release of baryons from the {\it drag} of CMB photons, occurring at a redshift ($\zd$). The scale $r_{\rm d}$ is thus the distance that sound waves in the photon-baryon fluid were able to travel all the way from the big bang, slightly after the time of recombination, to the drag epoch, given by
\begin{equation}
    \rd = \int_{z_{\rm d}}^\infty \frac{c_{\rm s}(z)}{H(z)} dz~\,,
    \label{eqn:rd}
\end{equation}    
 where $c_{\rm s}(z)$ is the speed of sound waves in the fluid, and $z_{\rm d} \approx 1060$ is the redshift at which photons and baryons decouple \cite{Planck-2018-cosmology}. 
 The BAO measurements are sensitive to the distance in the direction transverse to the line-of-sight, corresponding to the comoving angular diameter distance 
 \begin{equation}
    \DM(z) = \frac{c}{H_0} \int_{0}^{z} dz^\prime \frac{H_0}{H(z^\prime)}\,.
  \label{eqn:DM}
\end{equation}
BAO also measure the comoving distance along the line-of-sight, which is directly related to the expansion rate as
\begin{equation}
    \DH(z) = \frac{c}{H(z)}~.
    \label{eqn:DH}
\end{equation}
However, as described in \cite{DESI.DR2.BAO.cosmo}, some DESI BAO measurements are isotropic, as in the case of the BGS tracer. Hence, we also make use of the spherically-averaged distance $\DV(z)$ that quantifies the average of the distances measured along, and perpendicular to, the line-of-sight to the observer \cite{2005ApJ...633..560E}, and is given by 
\begin{equation}
\DV(z) = \left (z\DM(z)^2 D_H(z)\right )^{1/3}.
\label{eq:DV}
\end{equation}
Since these measurements are relative to the sound horizon $r_{\rm d}$, which sets the BAO scale, the directly constrained quantities are the ratios $\DM/\rd, \ \DH/\rd$, and $\DV/\rd$. With this, we can now define the primary dataset used for our searches of dynamical dark energy, based on the latest DESI data:

\begin{itemize}
    \item \textit{Baryon acoustic oscillations (BAO):} 
    We use the BAO distance measurements from DESI DR2, as detailed in Table III in \cite{DESI.DR2.BAO.cosmo}. In particular, for the BGS tracer, we use measurements of $\DVrd$ providing compressed low redshift information from the range $0.1<z<0.4$. For the rest of DESI tracers, we use the
    BAO distance measurements of $\DMrd$ and $\DHrd$. Explicitly, we use two LRG bins in the ranges $0.4<z<0.6$ and $0.6<z<0.8$, a combined tracer measurement for LRG+ELG in the range $0.8<z<1.1$, a measurement spanning $1.1<z<1.6$ for the ELG tracer and the QSO in the range $0.8<z<2.1$. The systematics tests associated with the BAO measurements from galaxy and quasar clustering are presented in \cite{Y3.clust-s1.Andrade.2025}. We also include the Ly$\alpha$ measurements in $1.8<z<4.2$, which provides our highest redshift data-point. This measurement is described in detail in \cite{DESI.DR2.BAO.lya} (see also \cite{Y3.lya-s1.Casas.2025} for validation tests and \cite{Y3.lya-s2.Brodzeller.2025} for specific catalog details). We refer to this whole dataset, encompassing information from redshift $0.1$ to $4.2$, split into seven main samples, as ``DESI''. 
\end{itemize}

We now proceed to define the cosmological datasets that we use, in combination with DESI, to obtain constraints on cosmological parameters. The cosmological probes and specific external datasets used in our analysis are: 

\begin{itemize}
     
    \item \textit{Supernovae Ia (SNe~Ia):} We combine DESI data with either of the following three SNe~Ia datasets, namely PantheonPlus, Union3, and DESY5.  The PantheonPlus \cite{Brout:2022} dataset comprises 1550 spectroscopically-confirmed SNe~Ia in the redshift range $0.001<z<2.26$.  The Union3 compilation \cite{Rubin:2023} has 2087 SNe~Ia in the redshift range  $0.01<z<2.26$, 1363 of which are common to PantheonPlus, though the analysis methodologies are substantially different. Finally, the DESY5  dataset \cite{DES:2024tys} is a sample of 1635 photometrically-classified SNe~Ia with redshifts in the range $0.1<z<1.13$, complemented by 194 historical low-redshift SNe~Ia (which are also present in the PantheonPlus sample) spanning $0.025<z<0.1$. 
    
    \item \textit{Cosmic microwave background (CMB):} We include temperature and polarization measurements of the CMB from the Planck satellite \cite{Planck-2018-overview}. In particular, we use the high-$\ell$ TTTEEE likelihood (\texttt{planck\_NPIPE\_highl\_CamSpec.TTTEEE}), together with low-$\ell$ TT (\texttt{planck\_2018\_lowl.TT}) and low-$\ell$ EE (\texttt{planck\_2018\_lowl.EE}) \cite{Aghanim:2019ame, Efstathiou:2021}, as implemented in \texttt{Cobaya}~\cite{Torrado:2021}. Additionally, we combine temperature and polarization anisotropies with CMB lensing measurements from the combination of NPIPE PR4 from Planck \cite{Carron:2022eyg,Rosenberg:2022} and the Atacama Cosmology Telescope (ACT) DR6 \citep{Madhavacheril:ACT-DR6,ACT:2023oei}.

    \item \textit{Compressed CMB}: We use the Gaussian correlated prior over $\omega_{\rm b} \equiv \Omega_{\rm b} h^2$, $\omega_{\rm bc} \equiv \Omega_{\rm b} h^2  + \Omega_ch^2$ and $\theta_*$ as defined in~\cite{DESI.DR2.BAO.cosmo}. Here, the angular acoustic scale $\theta_*$ adds extra geometrical information from the CMB, while $\omega_{\rm b}$ and $\omega_{\rm bc}$\footnote{Note that, as pointed out in our discussion about neutrinos, they do not contribute to the matter content of the Universe during recombination, and therefore we use $\omega_{\rm bc}$ explicitly instead of $\omega_{\rm m}$.} serve to set the sound horizon $r_{\rm d}$ and calibrate our BAO measurements. 
    These CMB-based quantities capture most of the relevant information from the early CMB by marginalizing over contributions from late-time effects, such as the integrated Sachs–Wolfe (ISW) effect and CMB lensing, resulting in a robust CMB compression for testing late-time physics \cite{EarlyUniverseCompression}.
    In particular, we use these compressed measurements as a conservative alternative for constraining dark energy at the background level, thereby allowing for negative $\fde(z)$, as in \cref{sec:crossing,sec:bins}.
    For brevity, we refer to these as $(\theta_\ast,\omega_\mathrm{b},\omega_\mathrm{bc})_\mathrm{CMB}$.
\end{itemize}

In our analysis, we utilize Markov Chain Monte Carlo (MCMC) sampling to explore the parameter space using the Metropolis-Hastings algorithm \cite{LewisMCMC:2002,LewisMCMC:2013} as implemented in \texttt{Cobaya}~\cite{Torrado:2020dgo, NealDragging:2005}.
For the alternate parametrizations, non-parametric methods, and DE classes, we adopt priors similar to \cite{DESI2024.VI.KP7A}, with exact specifications presented in \cref{tab:priors}, and have modified the Boltzmann solver \texttt{camb} \cite{LewisCAMB:2000,HowlettCAMB:2012}, incorporating a generalized equation of state for dark energy for the theoretical prediction of observables. We employ the parametrized post-Friedmann (PPF) framework \cite{Hu:2007,Fang:2008sn} to compute cosmological perturbations for the time-dependent equation of state $w(a)$, where $a$ is the scale factor, which permits transitions across the phantom divide at  $w = -1$. Additionally, we use custom theory code in \texttt{Cobaya} for the analysis of $f_{\rm DE}(z)$ binning and crossing statistics. For quintessence models, we have a modified version of the \texttt{class} \cite{Lesgourgues:2011re,Blas:2011rf} integrated into our inference pipeline. We switched to the \texttt{Recfast} option for recombination as it does not assume anything about the equation of state. We assume one massive and two massless neutrino species a with $\sum m_\nu=0.06 ~\rm eV$ and $N_{\rm eff} = 3.044$. For the SNe Ia likelihoods (PantheonPlus, Union3, and DESY5), we analytically marginalize over the absolute magnitude $M_B$. 
For clarity of presentation, we utilize Union3 in the figures as a conservative result, as it has larger uncertainties compared to the PantheonPlus and DESY5 datasets. Nevertheless, we will also discuss constraints derived from other supernova datasets wherever they are relevant to our analysis.
Finally, $\Delta\chi^2$ is defined with respect to the $\Lambda$CDM best fit. For the calculation of the best fit points themselves, we start with the maximum a posteriori (MAP) points from the four chains produced during the MCMC sampling, and make use of the \texttt{iminuit} \cite{iminuit} minimizer. Thus, the quantity used in model comparisons is more precisely\footnote{In the text $\chisq$ is used in place of $\chisq_{\rm MAP}$, for convenience.} $\dchisqMAP\equiv-2\Delta\ln \mathcal{L}$, which is the difference in the log posterior values at the calculated maximum posterior points, scaled by $-2$. Since the posterior depends on the product of the likelihood and priors, we also take into account the ratios of different model priors to ensure that there is no additional penalty in the $\dchisqMAP$ from comparing two models.

\section{Overview of the \tpdf{$w_0w_a$CDM} results} 
\label{sec:Overview}

\begin{figure}
    \centering
    \includegraphics[width=\columnwidth]{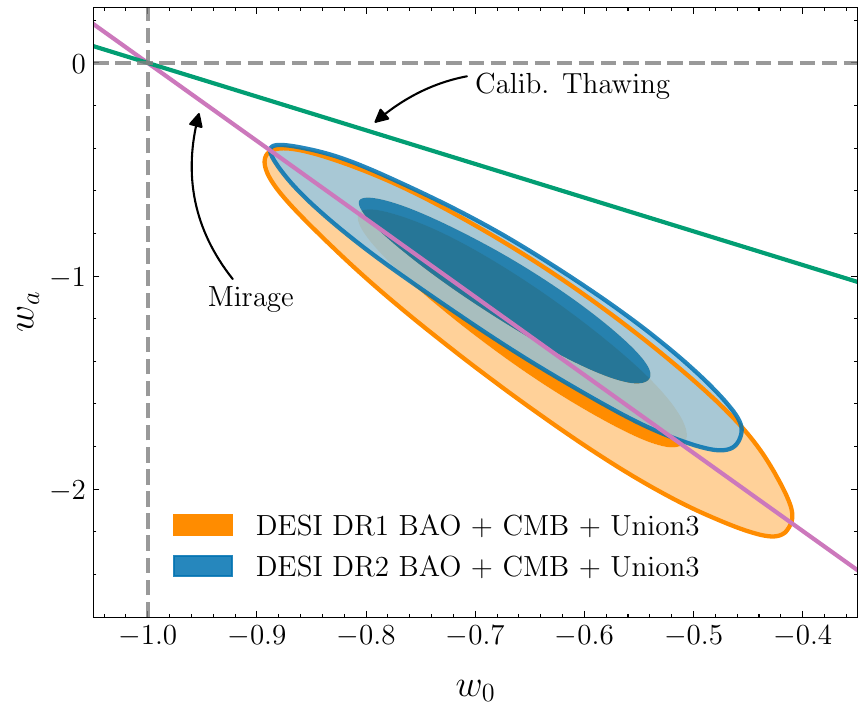}
    \caption{
 Constraints on the parameters $w_0w_a$ from DESI BAO DR2, CMB, and Union3 are illustrated in  blue, while the corresponding combination with DESI BAO DR1 is shown in orange. The green line indicates the degeneracy direction associated with calibrated thawing (see \cref{sec:thaw}), while the purple line denotes the ``mirage" direction (discussed in \cref{sec:mirage}) which closely follows the degeneracy direction of the $w_0w_a$ contours.}
    \label{fig:w0wa-plane}
\end{figure}

\begin{figure}
    \centering
    \includegraphics[width=1\columnwidth]{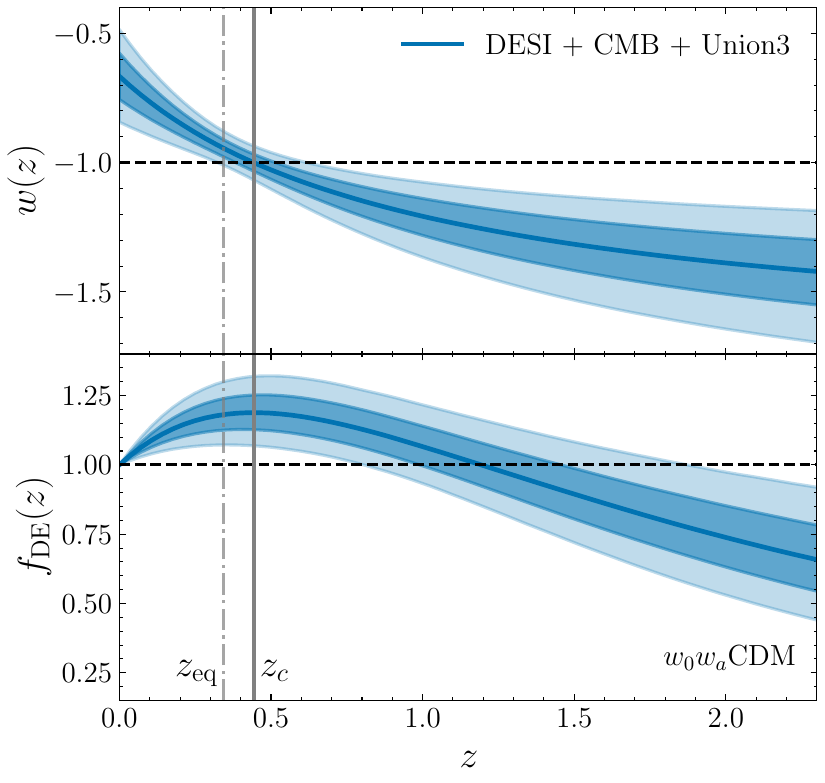}
    \caption{Equation of state parameter, $w(z)=P/\rho c^2$, and corresponding normalized dark energy density, $\fde (z) \equiv\rho_{\rm DE}(z)/\rho_{\rm DE,0}$, as a function of redshift using the $w_0w_a$ parametrization. The solid and dashed-dotted vertical lines indicate the phantom-crossing ($z_{\rm c}$) and dark energy-matter equality ($z_{\rm eq}$) redshifts, respectively. The horizontal dashed line represents $\Lambda$CDM. }
    \label{fig:w_fde_w0waCDM}
\end{figure}

We begin by summarizing the dark energy main findings of the DESI DR2 BAO key paper \cite{DESI.DR2.BAO.cosmo}, assuming the $w_0w_a$ parametrization given in \cref{eq:w0wa}.  As an example, the marginalized constraints in the $w_0w_a$ plane are shown in \cref{fig:w0wa-plane} for the DESI+CMB+Union3 data combination, together with the constraints from DESI DR1 BAO with those obtained with DESI DR2 BAO, corresponding to one and three-years worth of data, respectively. 
The combined data favor the region $w_0 > -1$ and $w_a < 0$, away from a cosmological constant, implying that the equation of state was phantom-like ($w(z) < -1$) in the distant past and has since evolved to $w(z) > -1$ at present, as shown in the top panel of \cref{fig:w_fde_w0waCDM}. 
This preference was observed in previous DESI analyses \cite{DESI2024.VI.KP7A,DESI2024.V.KP5,DESI2024.VII.KP7B,DESI:2024aqx,DESI:2024kob} and persists even when allowing for variations in the spatial curvature ($\Omega_{\rm K}$) \cite{DESI2024.VI.KP7A}, modified gravity \cite{KP7s1-MG}, or modifications to the pre-recombination physics \cite{Poulin:2024ken}.

DESI DR2 BAO data show that the mean posterior distributions have shifted slightly toward the \lcdm-expected values, while the reduced uncertainties have marginally increased the statistical significance of the deviations from \lcdm \ to $2.8 - 4.2\sigma$ (with improvements in fit ranging from $-21.0\leq\Delta\chi^2\leq-10.7$), compared to $2.5- 3.9\sigma$ from DR1 \cite{DESI.DR2.BAO.cosmo,DESI2024.VI.KP7A}. Similar conclusions follow when using the weighted posterior average of log-likelihood, the Bayesian counterpart of $\dchisq$. For a detailed Bayesian model comparison, see \cref{sec:Model-Comparison}.
Interestingly, with the increased precision, the combined DESI+CMB data already suggest a $\simeq3\sigma$ deviation from \lcdm, independent of any SNe~Ia compilation, with similar conclusions drawn from the DESI+DESY3 ($3\times2\rm pt$) combination, though exhibiting a lower tension; see Fig. 14 in \cite{DESI.DR2.BAO.cosmo}.

Physically, a phantom equation-of-state ($w (z)<-1$) translates into an energy density that increases with the expansion ($d\rho_{\rm DE}/da>0$), before reaching its maximum (at $z_{\rm c}\simeq0.45$, in our case), when the equation of state crosses the phantom line ($w(z_{\rm c})=-1$), and $\rho_{\rm DE}$ starts diluting again as the universe expands. The mean redshift at which this transition occurs in the $w_0w_a$ parameterization is indicated by a solid vertical line in \cref{fig:w_fde_w0waCDM}.
We should note at this stage that the exact redshift at which this crossing happens depends on the dataset combination under consideration. 

These results may naively suggest a ``phantom crossing'' \cite{Caldwell:1999ew} at high redshifts. From a theoretical perspective, this so-called phantom crossing is challenging to accommodate within standard scalar-field models of dark energy that are minimally coupled to gravity, as these are constrained to satisfy \(-1 \leq w \leq 1\). In particular, within general relativity, a single-field dark energy component with \(w < -1\) would necessarily violate the null energy condition (NEC), given by \(\rho c^2 + P \geq 0\) \cite{Hawking:1973uf}. If confirmed, the phantom crossing would have profound implications for fundamental physics, as it would indicate a significantly more complex dark sector than traditionally assumed. However, it is important to emphasize at this stage that the $w_0$-$w_a$ parametrization is particularly effective at capturing the impact of various, possibly more fundamental, dark energy models on observables such as distances and the expansion history within $\sim 0.1 \%$ accuracy \cite{dePutter2008} and may fail to accurately approximate the true behavior of $w(z)$ itself, potentially leading to a spurious indication of phantom crossing. Thus, restricting our analyses to models satisfying $w>-1$ might artificially bias our inference. For more discussion on phantom crossing, see \cref{sec:phantom}.

\begin{figure}
    \centering
    \includegraphics[width=\columnwidth]{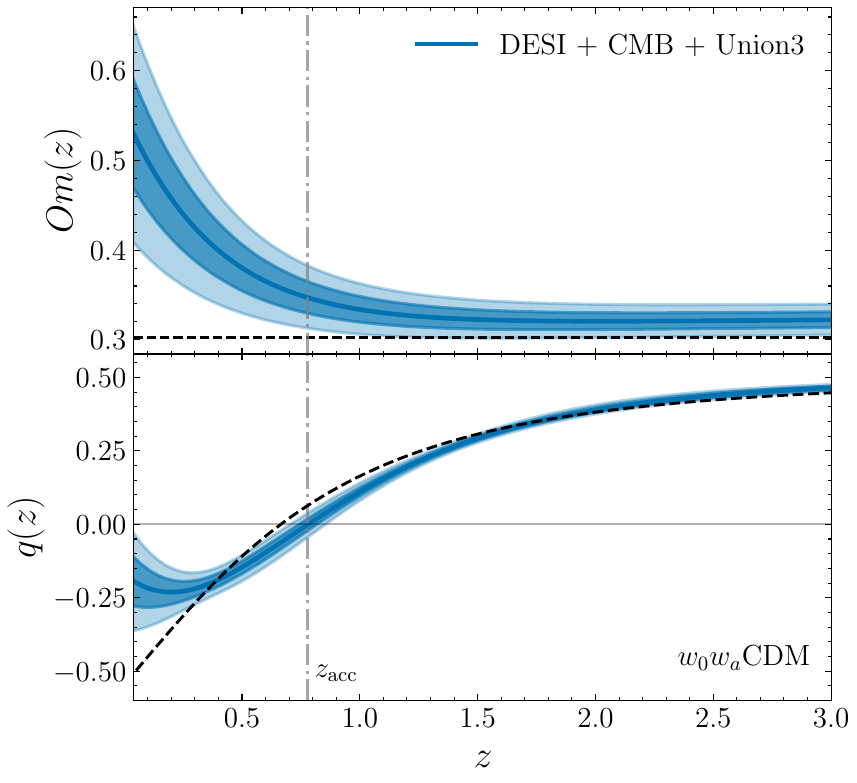}
    \caption{Evolution of the $Om(z)$ diagnostic and deceleration parameter, $q(z)$, as a function of redshift in the $w_0w_a$CDM model. The solid blue lines correspond to the median, $68\%$, and $95\%$ confidence levels obtained from the DESI+CMB+Union3 combination. The black dashed line depicts the best-fit \lcdm\ for the same data combination. The gray vertical line shows the redshift ($z_{\rm acc}$) corresponding to the onset of cosmic acceleration ($\ddot{a}>0$).}
    \label{fig:Om-q-w0waCDM}
\end{figure}

Before extending our analysis beyond $w_0w_a$CDM, we introduce two key quantities that will be useful throughout this work. \cref{fig:Om-q-w0waCDM} presents the $Om(z)$ diagnostic \cite{Sahni:2008xx} and the deceleration parameter $q(z)$ for the $w_0w_a$CDM model, where 
\begin{equation}\label{eq:Omz}
  Om(z)\equiv\frac{h^2(z)-1}{(1+z)^3-1}~,
\end{equation}
and the deceleration parameter is given by
\begin{equation}\label{eq:q(z)}
    q(z)\equiv{-\frac{\ddot{a}a}{\dot{a}^2}}=\frac{d\ln H}{d\ln(1+z)}-1~.
\end{equation}
These two functions constitute a sensitive probe of new physics, as they are only sensitive to the `shape' of the (normalized) expansion history $h(z)=H(z)/H_0$. Thus, they are unaffected by the degeneracies that may exist between the dark energy and matter densities 
at the background level \cite{Wasserman:2002gb,PhysRevD.80.123001,PhysRevD.84.063519}. 
Indeed, one can readily see from \cref{eq:Omz} that the quantity $Om(z)$ is strictly constant and equal to present matter density  ($Om(z)\equiv \Om$) if dark energy is in the form of a cosmological constant. Thus, $Om(z)$ serves as a null-test of \lcdm, and any significant deviation from a constant value would indicate dynamics in the dark energy density.
The reconstructed $Om(z)$ in \cref{fig:Om-q-w0waCDM} shows a clear ($>2\sigma$) deviation from constancy in the range $0\lesssim z\lesssim0.5$, where the black dashed line represents the best-fit \lcdm\ value of $\Omega_m=0.302$. On the other hand, $q(z)$ tracks the logarithmic derivative of $h(z)$, rather than its shape, approaching a value of $0.5$ during matter domination. The reconstructed $q(z)$ suggests that the Universe's acceleration ($q<0$) began earlier in cosmic history ($z_{\rm acc}\simeq0.8$) than predicted by \lcdm\ ($z_{\rm acc}\simeq0.65$) with a slowing down of cosmic acceleration at recent times. These trends in $Om(z)$ and $q(z)$ were previously observed
with DESI DR1 data and persist with slightly more statistical significance in DESI DR2.

\section{Parameterizing Dark Energy}\label{sec:params}

To more closely explore the possible dynamical nature of dark energy, we now turn to parameterizations of either the equation of state $w(z)$, or energy density $\rho_{\rm DE}(z)$. 
Since different parameterizations can lead to differences in the inferred evolution of dark energy, it is crucial to explore multiple forms to assess the robustness of any detected deviation from a cosmological constant. We examine various two-parameter functional forms as alternatives to $w_0w_a$CDM. In addition, we increase the degrees of freedom available to $w(z)$, to probe the trends present in the data. While the parameterizations investigated here are not necessarily tied to a specific physical model, they cover distinct functional spaces, helping to ensure the results are not driven by the choice of parametrization.

\subsection{Alternative \tpdf{$w(z)$} parameterizations }\label{sec:alt-par}

\begin{figure}
    \centering
    \includegraphics[width=\columnwidth]{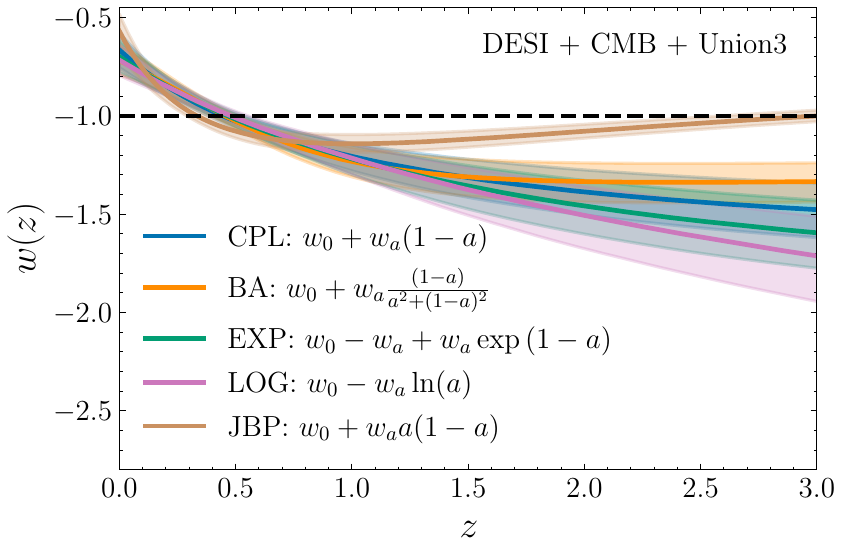}
    \caption{ 
The dark energy equation of state $w(z)$ for alternative $w_0w_a$ parameterizations \textemdash BA, EXP, JBP, and LOG \textemdash is illustrated alongside the standard CPL form (shown in blue). The constraints integrate data from DESI, Union3 SNe, and CMB observations, with shaded regions representing $1\sigma$ uncertainty bands.
    All parameterizations except JBP exhibit consistent phantom crossing 
    near $z \sim 0.5$ and provide a similarly good fit to the data.
    }
    \label{fig:wz_alternative}
\end{figure}

In this section, we explore four alternative parameterizations from the literature (see \cite{Giare:2024gpk,Wolf:2025jlc} for the equivalent DR1 results) that, like \( w_0w_a \)CDM, introduce two additional parameters: the present-day equation of state \( w_0 \) and an evolution parameter \( w_a \), but with different functional forms.  

\Cref{fig:wz_alternative} presents constraints on these alternative models as defined in \Cref{tab:w0wa_alternative}, i.e.: Barboza-Alcaniz (BA) \cite{Barboza2008,Efstathiou1999}, exponential (EXP\footnote{In the numerical implementation we truncate at $3^{\rm rd}$ order in Taylor expansion.}) \cite{Dimakis2016,Pan2020}, logarithmic (LOG) \cite{Efstathiou1999} and Jassal-Bagla-Padmanabhan (JBP) \cite{Jassal2005}, alongside the Chevallier-Polarski-Linder (CPL) in blue for comparison. The shaded bands, representing 1$\sigma$ uncertainties, are derived from a combination of DESI, CMB, and Union3. With the exception of JBP, all models exhibit similar low-redshift behavior, including a phantom crossing near $ z \sim 0.5 $. In \Cref{tab:w0wa_alternative}, we present the alternative functional forms of $w(a)$ and $ \Delta \chi^2 $ values relative to $\Lambda$CDM, showing that BA, CPL, LOG, and EXP provide statistically comparable fits to the data. The functional form of the JBP parametrization, which forces it to assume identical early- and late-time behavior, results in a slightly poorer fit. These findings confirm that constraints from CPL are broadly representative of the alternative $ w_0w_a $ models considered, with no significant improvement observed for any alternative form. This suggests that current data lacks the sensitivity to distinguish between these parameterizations at $ z > 2 $, a conclusion that remains unchanged across different SNe~Ia datasets.

\begin{table}
    \centering
    \caption{$\dchisqMAP$ values relative to $\Lambda$CDM for alternative $w_0 w_a$ parametrization using DESI + CMB + Union3.}
    \label{tab:w0wa_alternative}
    \begin{ruledtabular}
    \begin{tabular}{lcc}
        Param. & Functional Form & $\Delta\chi^2$ \\ \hline
        BA   & $w_0 + w_a \frac{1-a}{a^2+(1-a)^2}$ & $-17.3$ \\
        EXP  & $(w_0 - w_a) + w_a \exp(1 - a)$       & $-17.5$ \\
        LOG  & $w_0 - w_a \ln a$                      & $-17.6$ \\
        JBP  & $w_0 + w_a\,a(1-a)$                     & $-13.6$ \\
        CPL  & $w_0 + w_a (1-a)$                       & $-17.4$ \\
    \end{tabular}
    \end{ruledtabular}
\end{table}

\subsection{Crossing Statistics}
\label{sec:crossing}

Rather than exploring different redshift evolutions for $w(z)$, one can instead gauge the impact of introducing additional degrees of freedom in the 
DE characteristics.
Following the methodology detailed in \cite{DESI:2024aqx}, we expand the equation of state of dark energy $w(z)$ in terms of Chebyshev polynomials (see also \cite{shafieloo2011crossing,Shafieloo_2012,Shafieloo_2012b,Haude:2019qms}), 
\begin{equation}\label{eq:crossing_w}
    w(z) = -\sum_{i=0}^{N} C_i \, T_i(x)~,
\end{equation}
where $C_i$ are free coefficients, and $T_i(x)$\footnote{Note that the redshift interval relevant for observations $z\in[0,3.5]$ is mapped to $x\in[-1,1]$, where the Chebyshev polynomials are defined.} are Chebyshev polynomials of the first kind, forming a complete basis for continuous functions in the large-$N$ limit, although $N\simeq3$ is generally sufficient to capture smooth functions. We note that \lcdm\ is recovered for $C_0=1$ and $C_{i>0}=0$.  Alternatively, one may want to work with the normalized dark energy density $\fde(z)$ instead
\begin{equation}\label{eq:crossing_fde}
        \fde(z) = \sum_{i=0}^{N} C_i \, T_i(x)~.
\end{equation}
Expanding in $\fde(z)$ offers the advantage of allowing the (effective) energy density $\rho_{\rm DE}(z)$ to change sign, thereby encompassing a broader class of models  \cite{Grande:2006nn,Vazquez:2012ag,Visinelli:2019qqu,Calder_n_2021}, including modified gravity scenarios \cite{Chiba:1999ka,Sahni_2003,Bauer_2010,Boisseau:2015hqa} and complex dark sector interactions that are difficult to capture with a parametrized $w(a)$. We note that the expansion in \cref{eq:crossing_fde} has one less degree of freedom relative to that of \cref{eq:crossing_w}, as all samples must satisfy $\fde(z=0)=1$.

\begin{figure}
    \centering
    \includegraphics[width=\columnwidth]{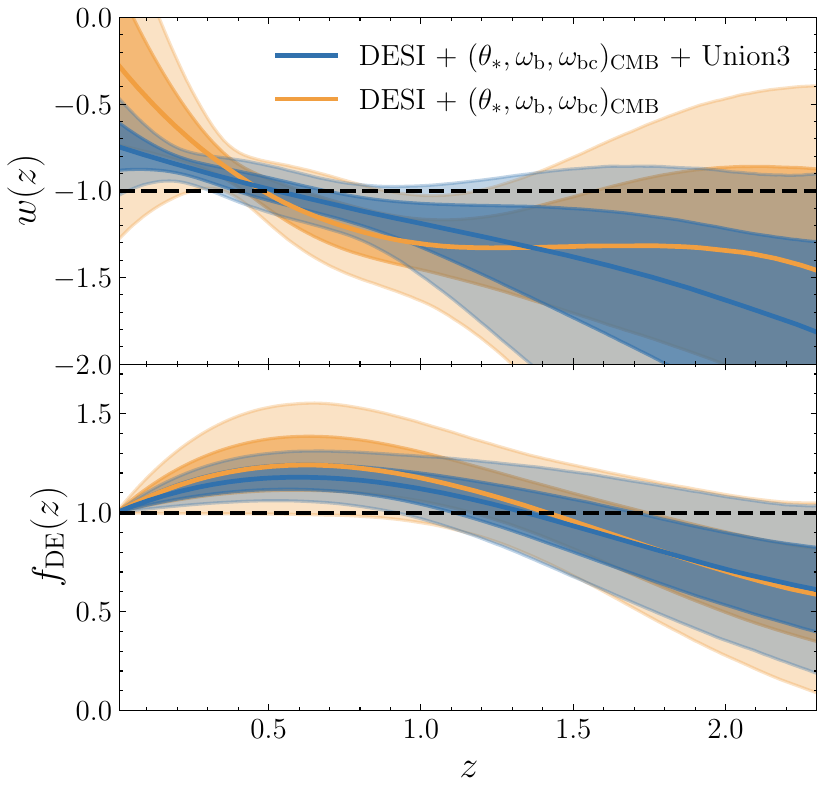}
    \caption{Reconstructions of $w(z)$ and $\fde(z)$ using \cref{eq:crossing_w,eq:crossing_fde} with $N=3$ for the DESI + $(\theta_\ast,\omega_\mathrm{b},\omega_\mathrm{bc})_\mathrm{CMB}$ data combination, with and without the inclusion of Union3. The solid lines correspond to the median, $68\%$, and $95\%$ confidence levels around it. The main reconstructed behavior of DE is in excellent agreement with the different model-agnostic approaches explored in this paper. This confirms that the trend is not driven by the choice of parametrization, and adds to the robustness of the $w_0w_a\rm CDM$ results presented in the main key paper \cite{DESI.DR2.BAO.cosmo}. The horizontal dashed line represents $\Lambda$CDM. }
    \label{fig:crossing_w_fde}
\end{figure}

The top panel of \cref{fig:crossing_w_fde} shows the reconstructed $w(z)$ for the DESI+$(\theta_\ast,\omega_\mathrm{b},\omega_\mathrm{bc})_\mathrm{CMB}$ 
combination, with (blue) and without  (orange) SNe Ia data from the Union3  compilation. The bottom panels show the reconstructed $\fde(z)$. 
It is noteworthy that not only do the expansions in $ w(z)$ and  $\fde(z)$ yield similar behaviors independently of SNe Ia data, but they also agree with the main results obtained using the $w_0w_a$CDM parametrization \cite{DESI.DR2.BAO.cosmo}, as shown in \cref{fig:w_fde_w0waCDM}. This consistency further strengthens the robustness of the results.

\begin{figure}
    \centering
    \includegraphics[width=\columnwidth]{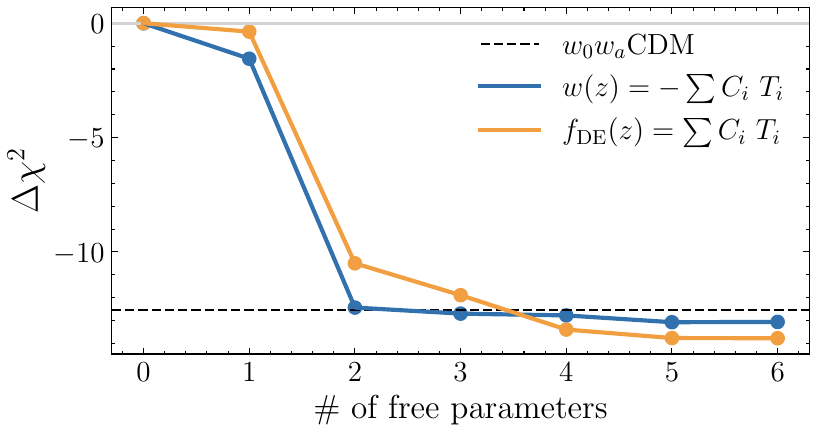}
    \caption{Improvement in fit with respect to \lcdm, as quantified by $\Delta\chi^2=\chi^2-\chi^2_{\Lambda\rm CDM}$, as a function of the number of free parameters introduced in the expansion, for the DESI + $(\theta_\ast,\omega_\mathrm{b},\omega_\mathrm{bc})_\mathrm{CMB}$ + Union3 combination.}
    \label{fig:delta_chi2_N_crossing}
\end{figure}

While these results perfectly align with the $w_0w_a$CDM results, the expansions in \cref{eq:crossing_w,eq:crossing_fde} offer greater flexibility, enabling it to capture features in the evolution of dark energy beyond the linear parametrization given by~\cref{eq:w0wa}.
Despite this additional flexibility—and also confirmed by our independent analyses—the combined data favor a smooth evolution, well described by $w_0w_a$CDM within the probed low-redshift range. The improvement in fit, quantified by $\Delta\chi^2$, is shown as a function of the number of free parameters in \cref{fig:delta_chi2_N_crossing}. A two-parameter expansion in $\fde(z)/w(z)$ captures the main trends in the data, as already noted in \cite{Linder:2005ne}.
Introducing additional degrees of freedom does not significantly improve the fit to the combined data and would be disfavored from a model comparison perspective, as the added complexity is not justified by the data. 
We note that due to the complications that can arise in the treatment of perturbations when allowing for $\rho_{\rm DE}<0$, this part of the analysis is restricted to the ``compressed'' 
CMB information, denoted as $(\theta_\ast,\omega_\mathrm{b},\omega_\mathrm{bc})_\mathrm{CMB}$, rather than the full Planck likelihood. We have verified that $(\theta_\ast,\omega_\mathrm{b},\omega_\mathrm{bc})_\mathrm{CMB}$, as described in \cref{sec:data}, yields almost identical constraints as those using the primary CMB anisotropies.

While all the parametric models we tested above suggest a phantom crossing, the exact redshift at which this happens depends on the chosen parametrization, as seen from \cref{fig:crossing_w_fde} which suggests a slightly higher value for $z_{\rm c} \simeq 0.5$ than $w_0w_a$CDM. This variation\textemdash although not statistically significant\textemdash is expected due to the inherent limitations of parametric fitting, as each functional form has a restricted degree of flexibility. 

\section{Non-Parametric Methods}\label{sec:non-params}

In contrast to the techniques explored in \cref{sec:params}, non-parametric techniques focus on determining the true function of quantities such as $h(z)$, $\fde(z)$, and $w(z)$ from observational data, rather than merely estimating the parameters of a pre-specified form for $w(z)$. We are not interested in model comparison here \textit{per se}, but rather the robustness of the observed trends in the data under different non-parametric reconstruction techniques.

We explore two techniques: binning and Gaussian process regression. We have also tested the cosmographic expansion up to $\mathcal{O}((t-t_0)^5)$  where $t$ is the lookback time, and $t_0$ denotes the current epoch \cite{Muthukrishna:2016evq,DES:2024fdw}. However, we do not present those results here since they did not pass validation tests with the full DESI data and require a redshift cut-off to make unbiased inference.

\subsection{Binning}
\label{sec:bins}
Binning is a technique widely used in cosmology that allows for comparison of different redshift intervals, without the assumption of a specific functional form; see ~\cite{Tegmark_1997,Huterer:2000mj,Huterer_2005,Crittenden:2005wj,Simpson:2006bd,dePutter2008,Garcia-Quintero:2020bac,Planck:2015bue,Zhao:2017cud,Raveri:2021dbu,Pogosian:2021mcs,Bansal:2025ipo} for some examples.
Here we focus on binning the equation of state of dark energy, $w(z)$ and the dark energy density, $\fde(z)$, permitting localized analyses of the behavior motivated by the data. The additional degrees of freedom introduced make it possible to probe for potential variations or trends in $w(z)$ across redshifts, which may help to identify deviations from the standard $\Lambda$CDM model.

In this section, we supplement the 3 uniform redshift bin scheme for the dark energy equation of state parameter from~\cite{DESI.DR2.BAO.cosmo} (see Figure 12 there) in order to assess the impact of different choices for the implementation.  
The binned function takes the general form
\be
w(z) = w_0 + \sum_{i=1}^{N}\frac{(w_i-w_{i-1})}{2}\left(1+\tanh\left(\frac{z-z_i}{s}\right)\right)\,,
\label{eq:binning}
\ee
where $w_i$ are the bin amplitude parameters, $N$ the number of bins and $s$ the smoothing scale\footnote{For this analysis $s=0.02$ is chosen, corresponding to less than 1\% variation in the bin amplitude over the range $\Delta z = 0.01$ on either side of the redshift bin edge.}, which controls the sharpness of the transitions around the edges $z_i$ between bins. We assume no prior correlation between bins.

Several different additional schemes for $w(z)$ were tested, including logarithmic binning, binning aligned with the redshifts of the tracer types, and various uneven binning approaches across the constrained redshift interval. However, for clarity, we present results only for schemes with uniform redshift bins between $z=0$ and $z=2.1$, as results do not change qualitatively across the different binning schemes.
We consider the combination of DESI, Union3 SNe~Ia, and CMB. In the case of $\fde(z)$, the compressed CMB is used to avoid the computational complexity associated with correctly modifying the behavior of dark energy in a Boltzmann solver to account for $\fde<0$, while also constraining the parameters exclusively with early CMB information. 

\begin{figure}
    \centering
    \includegraphics[width=1\columnwidth]{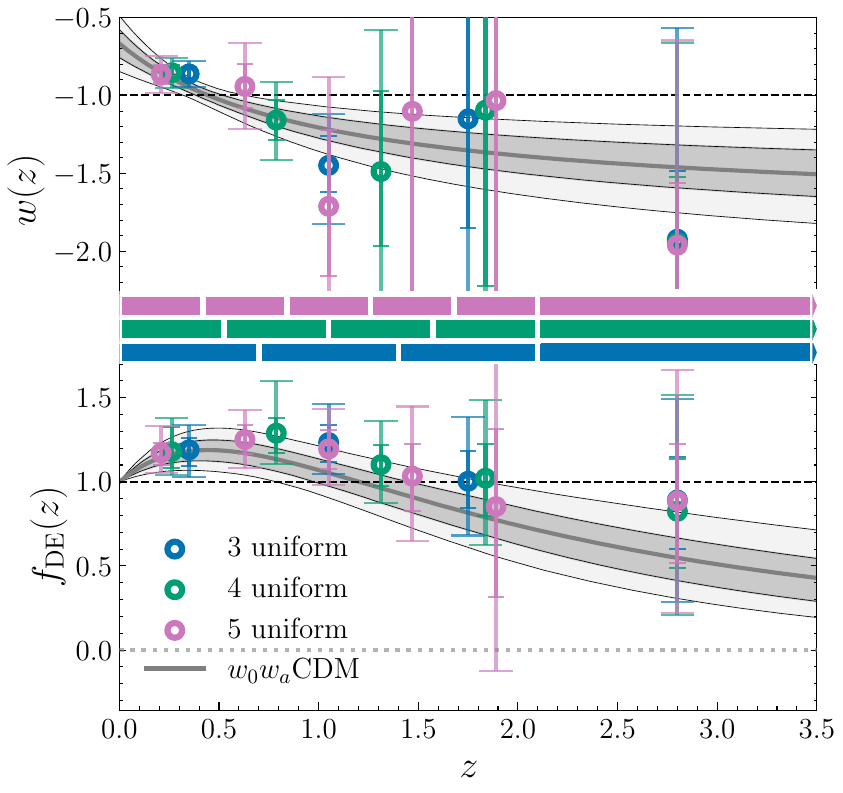}
    \caption{Median bin amplitudes with $1\sigma$ and $2\sigma$ error bars for $w(z)$ (upper panel), using DESI+${\rm CMB}$+Union3, and $\fde(z)$ (lower panel), using DESI+$(\theta_\ast,\omega_\mathrm{b},\omega_\mathrm{bc})_\mathrm{CMB}$+Union3. Results are shown for 3 schemes, with an increasing number of uniform bins in the range $0<z<2.1$. The redshift intervals of the bins in these different schemes are shown in the corresponding colored bars between the two panels. For comparison, the median, $1\sigma$, and $2\sigma$ contours of the $w_0w_a$CDM parametrization are plotted on the same axes, along with $\Lambda$CDM expectation in dashed lines. The $\fde=0$ line is plotted as a dotted line.  
    }
    \label{fig:fde_bin_comp}
\end{figure}

\cref{fig:fde_bin_comp} (upper panel) shows the median values of $w_i$, with $1\sigma$ and $2\sigma$ error bars, positioned at the center of their respective bins' redshift intervals. These intervals are shown in the same colors in between the panels. The highest redshift interval effectively extends to high redshifts, with the corresponding amplitude positioned at $z=2.8$ merely for convenience.
The constrained amplitudes for overlapping bins between different schemes are all within $\sim1\sigma$ of one another. Superimposing with the median (dashed grey line) and $1\sigma$, $2\sigma$ confidence levels of the corresponding $w_0w_a$CDM result, we see that the behavior recovered by each different scheme is in good general agreement, with median points on either side of the line $w(z)=-1$.
The data provide the tightest constraints in the lowest redshift bin, where they prefer a $w(z)$ that is more than $3\sigma$ away from $\Lambda$CDM value of $-1$, whereas the higher redshift bin amplitudes remain, at most, within $2\sigma$ of $\Lambda$CDM. 
The question of an actual crossing is more subtle, since it would have to occur at the edge between two adjacent bins, meaning that it would depend non-trivially on the number of bins and their chosen centers, and not make for a very robust `measurement'. 

To allow explicit exploration of the region of parameter space with negative $\fde(z)$, which is excluded when binning with the amplitude of $w(z)$, we also test additional binning schemes where the bin variables are instead associated with the amplitude of $\fde(z)$. \cref{fig:fde_bin_comp} (lower panel) shows the same effective behavior between individual binning schemes, with good agreement to the $w_0w_a$ curves, indicating a turn-over somewhere in the region of $0.5<z<1.0$ and $\fde(z)>0$ at around $2\sigma$ for most of the bins. The uncertainties increase with redshift, becoming progressively less Gaussian, with longer tails extending towards more negative $w$ values. Lastly, we note that the amplitudes in adjacent bins exhibit mild correlations, weakening with increasing redshift. 

To decorrelate the bins and get additional insights into the contributions from different redshift intervals, we also perform a principal component analysis. Principal component analysis (PCA) is effectively a transformation that provides a new basis in which the new coefficients $q_i$, corresponding to the bin amplitude parameters, are decorrelated. There are, in general, infinitely many such decorrelated bases, but only one that is orthogonal. We may obtain it simply by finding the eigenvector basis that diagonalizes $C^{-1}$, the inverse covariance matrix of the bin amplitude parameters $w_i$, calculated by MCMC sampling~\cite{Huterer_2003}. 
See \cref{sec:BinApp} for additional details.

\begin{figure}
    \centering
    \includegraphics[width=1\columnwidth]{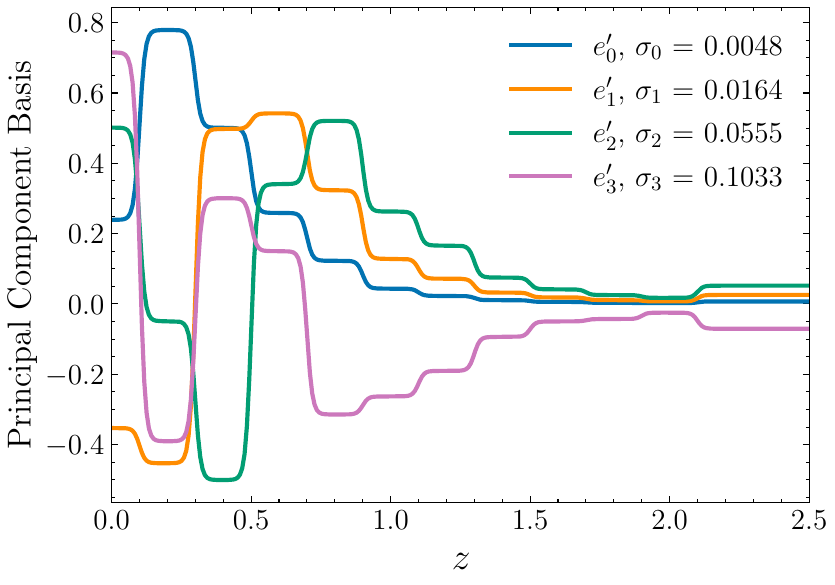}
    \caption{The four principal components with the largest eigenvalues, from a scheme comprising 10 uniform bins between $z=0.1$ and $z=2.1$, constrained using DESI+${\rm CMB}$+Union3. The first component has an uncertainty (inverse eigenvalue), $\sigma_0$, at least $20\times$ smaller than any of the others, and the first two components are seen to be relatively well localized at low redshift.}
    \label{fig:PCA_bin}
\end{figure}

We divide the equation of state parameter into 10 uniform bins of fixed amplitude between $z=0.1$ and $z=2.1$, with two additional free parameters, one each for the amplitude on either side. The covariance matrix of the resulting bin parameters $w_i, i=0,1,...,11$ with DESI+CMB+Union3 is used to determine the eigenvector basis.  
The basis functions, or principal components, corresponding to the 4 largest eigenvalues are presented in~\cref{fig:PCA_bin}, along with the corresponding errors (obtained as square roots of inverse eigenvalues)

The largest principal component is well localized in $z$, peaking in the range $0.1<z\leq0.3$, while the second-largest component is mostly positive and peaks in the interval $0.3<z\leq0.5$. The remaining components show increasingly more pronounced oscillatory behavior, with at least $20\times$ the uncertainty of the first bin, $\sigma_0$.

Overall, the binning results from different schemes are in good general agreement. The crossing of phantom divide line by $w(z)$, and turn-over in $\fde(z)$ followed by a decreasing trend towards higher redshifts, found in the other analyses \cite{DESI2024.VI.KP7A,DESI:2024aqx,Reboucas:2024smm,Pang:2024qyh} are consistent with these results. Even so, the approach has its limitations. While it is well-suited to testing deviation from a constant function, capturing more complicated behaviors requires additional degrees of freedom, which increases the level of uncertainty~\cite{Huterer_2003,Crittenden:2005wj,Simpson:2006bd,dePutter2008,Zhao:2017cud}. In particular, though the data seem to be consistent with a phantom crossing, it is difficult to draw strong conclusions about the specific redshift where a phantom crossing of $w(z)$, or turn-over in $\fde(z)$, might occur. The limitations present in this approach make it important to understand what kind of biases may be introduced by the implementation. In~\cref{sec:mock_val}, we perform some tests on simulated data in an attempt to address this. 

\subsection{Gaussian Process Regression}\label{sec:gp}

\begin{figure*}
    \centering
    \includegraphics[width=1\linewidth]{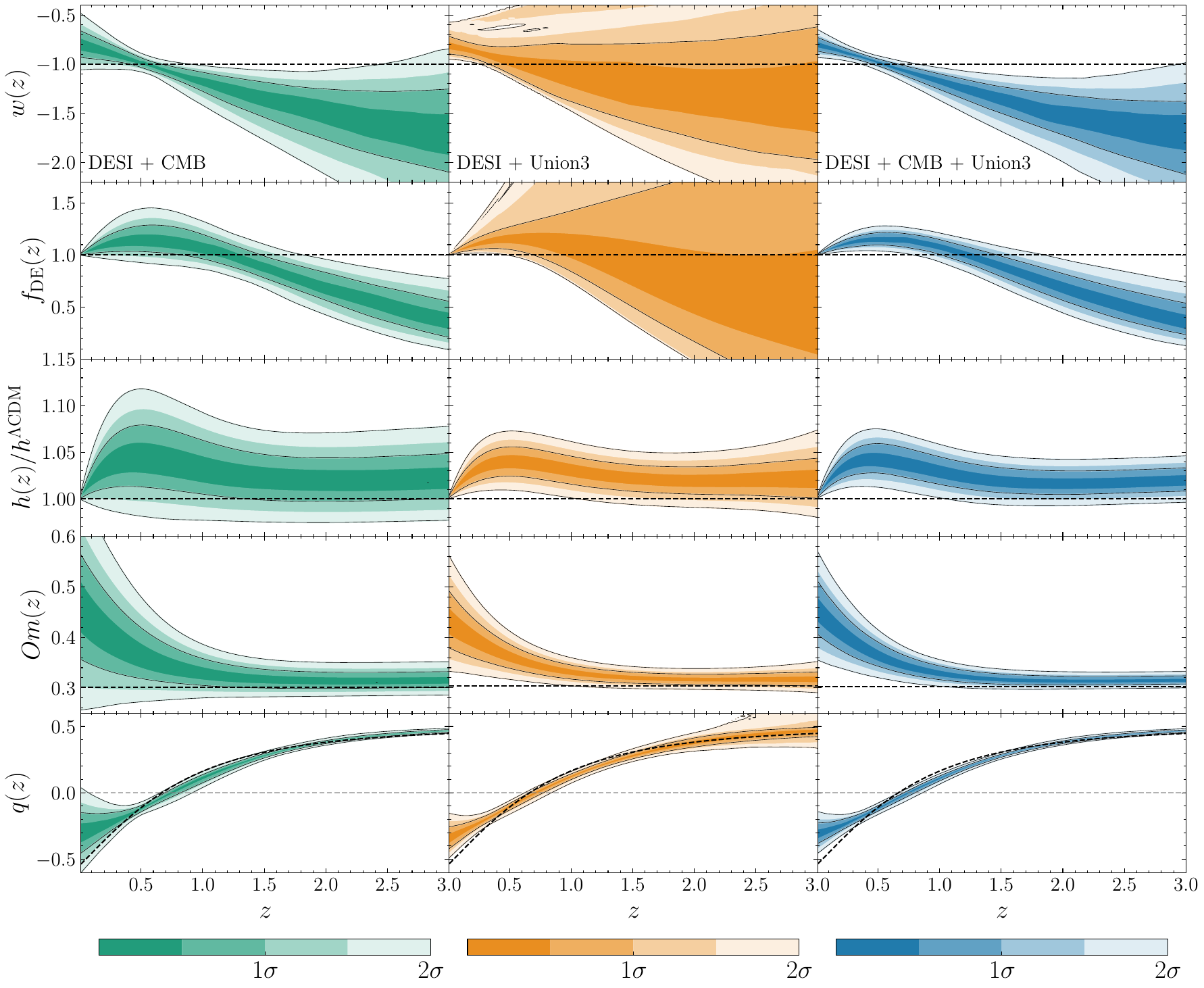}
    \caption{GP reconstructions of  $w(z)$  for the DESI + CMB (green), DESI + Union3 (orange), and DESI + CMB + Union3  (blue) combinations. The rows depict the redshift evolution of cosmological quantities: equation of state $w(z)$, normalized dark energy density  $f_{\rm{DE}}(z)$, $h(z)/h_{\Lambda\text{CDM}}(z)$, $Om(z)$ diagnostic, and deceleration parameter $q(z)$. The shaded bands obtained using \texttt{fgivenx} \cite{fgivenx} illustrate confidence intervals at various levels. The black dashed lines represent predictions from the  $\Lambda${CDM} model.
    }
    \label{fig:GP-w-grid}
\end{figure*}

\begin{figure}
    \centering
    \includegraphics[width=1\linewidth]{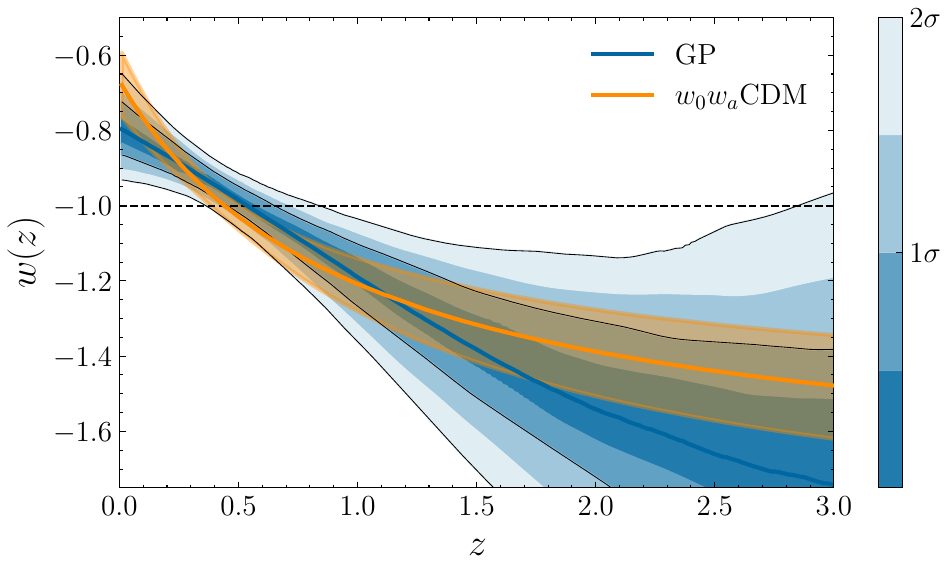}
    \caption{Comparison of GP reconstruction of dark energy equation of state $w(z)$ with the $ w_0w_a$ parametrization, utilizing data from DESI, CMB, and Union3. The GP reconstruction is shown in blue, accompanied by shaded $68\%$ and $95\%$ 
    confidence intervals. This is contrasted with the posterior predictions for the $w_0w_a$CDM model, depicted in orange along with 1$\sigma$ uncertainty. The black dashed line represents the standard $\Lambda$CDM prediction.}
    \label{fig:gp-cpl-comp}
\end{figure}

In this section, we discuss Gaussian processes, which can be thought of as a generalization of binning where the amplitudes at every redshift are sampled but are subject to some constraints (prior assumptions) on the form of the resulting functions. This allows for a complementary analysis with the possibility of improving the trade-off between flexibility and constraining power.

Gaussian Process (GP) regression \cite{rasmussen2006gaussian} is a powerful, non-parametric statistical tool widely used in various fields, including cosmology \cite{Holsclaw_2010,Holsclaw2010PRL,Shafieloo:2012ht}, to reconstruct smooth functions from noisy data without assuming a specific functional form (see e.g., \cite{Seikel_2012,Shafieloo2013,Keeley2019Implications,Belgacem:2019zzu,Keeley:2019hmw,Gerardi:2019obr,Mukherjee:2021kcu,Calderon:2022cfj, Calderon2023,Dinda:2024ktd, Mukherjee:2024ryz} for a non-exhaustive list). 
For the purpose of this work, GP can be thought of as a way of sampling the space of continuous functions in a non-parametric manner. This allows data-driven reconstructions of the quantities of interest for dark energy, namely $w(z)$ or $\fde(z)$ with minimal assumptions \cite{Holsclaw_2010,Joudaki:2017zhq,Calderon:2022cfj,Hwang:2022hla}. More specifically, at each point in parameter space, we draw a sample (realization) of $w$ from a multivariate Gaussian distribution, e.g., $w(z)\sim\mathcal{GP}(m(z)=-1,K)$ where $K$ is a given covariance function~\textemdash~known as kernel~\textemdash~encoding our prior assumptions about the smoothness of the reconstructed function. We further impose $w(z\geq z_{\rm max})=-1$ to recover a standard (\lcdm-like) expansion history at early times. We have chosen $z_{\rm max}=10$, after checking that this choice does not significantly alter our conclusions. This is implemented in a modified version of the Boltzmann solver \texttt{camb}, with more details on GP given in \cref{sec:GP-app}.

\cref{fig:GP-w-grid} illustrates the reconstructed dark energy properties using GP with various datasets: DESI + CMB (left), DESI + Union3 (middle), and DESI + Union3 + CMB (right). The top row presents the reconstructed $w(z)$, indicating deviations from $\Lambda $CDM at low redshift and hints of $w(z)$ crossing into the phantom regime around $z \approx 0.5$. The inclusion of CMB data (left and right panels) results in tighter constraints on $\Omega_{\rm m}$, which strengthen the significance of deviations from  $w = -1$, whereas the DESI + Union3 combination allows for lower values of $\Omega_{\rm m}$ and a broader range of variations $w(z)$.  The second row demonstrates a notable bump in the evolution of the dark energy density, while constraints derived without CMB allow for a wider variety of $w(z)$. The third row displays the normalized Hubble parameter  $h(z)/h_{\Lambda\text{CDM}}(z) $. The fourth row presents the $Om$ diagnostic, clearly showing the evolution as a function of  $z$, indicating a deviation from $\Lambda$. Lastly, the final row depicts the reconstructed deceleration parameter $q(z)$, which slightly exceeds the expectations of the $\Lambda$CDM model, suggesting a slowdown in the acceleration rate.

In \cref{fig:gp-cpl-comp}, we present a comparison of results obtained from GP reconstruction utilizing the $w_0w_a$ parametrization derived from DESI, CMB, and Union3 data. The GP reconstruction, illustrated in blue, aligns very well with the 1$\sigma$ posterior predictions of the $w_0w_a$CDM model. We would like to remind readers that the GP approach imposes a Gaussian prior distribution on $w(z)$, centered at the mean function which we explicitly choose to be $w = -1$, as represented by the black dotted line. This effectively places more prior weight on $\Lambda$ and any observed deviations from $w(z) = -1$ are largely driven by data. Finally, we would like to emphasize that although GP regression offers advantages over parametric methods, it is important to interpret the reconstructed $w(z)$ with caution. While flexible, the method may not fully capture certain behaviors of $w(z)$, as illustrated in \cref{sec:mock_val} using simulated data. Nevertheless, GP remains a valuable tool for assessing the dynamical nature of dark energy in a non-parametric manner.

\section{Implications for Dark Energy}
\label{sec:implications}

The various methods explored in~\cref{sec:params,sec:non-params} provide a flexible way to test deviations from \lcdm\ and ensure robust results without committing to a specific dark energy model. However, interpreting the deviations from a cosmological constant and understanding its implications for fundamental physics necessitates a deeper exploration of physically-motivated dark energy models. Rather than constraining specific models, we focus on different classes, characterized by their dynamics \cite{Caldwell:2005tm,Linder:2006sv,Cahn:2008gk} and inspired by theoretical considerations. 

\subsection{Thawing dark energy}\label{sec:thaw} 

The first class of models we consider is known as \textit{thawing} dark energy \cite{Caldwell:2005tm}. This class characterizes quintessence models \cite{PhysRevD.37.3406,Wetterich:1987fm,Ferreira:1997hj,Scherrer:2007pu}, in which a minimally coupled scalar field remains frozen at early times due to Hubble friction, effectively behaving like a cosmological constant with $ w = -1$. Only when the scalar field’s mass becomes comparable to the Hubble rate, $m_\varphi \sim H$, does the field begin to evolve dynamically, causing its equation of state to `thaw' away from $w = -1$ into the quintessence regime, $ w > -1 $. Note that there exists a second class of DE dynamics, referred to as the `freezing' class, where the field evolves towards a de Sitter state ($w=-1$) in the asymptotic future. Such dynamics are characterized by $w_a > 0$ and are not favored by observations. For a review on quintessence models, see e.g., \cite{Tsujikawa:2013fta,Martin:2008qp} 

This behavior is typical of pseudo-Nambu-Goldstone-Boson (PNGB) quintessence models \cite{Frieman:1995pm} and simple potentials such as $V \propto m^2\varphi^2 $ and  $V \propto \lambda\varphi^4 $, both of which are ubiquitous in high-energy physics \cite{Copeland:2006wr}. Interestingly, Ref.~\cite {dePutter2008} demonstrated that the phase-space dynamics of these models can be well-approximated using the $w_0w_a$ parametrization. 
Many thawing potentials map onto a narrow region in the $w_0w_a$ plane, approximately following the 
relation  
\begin{equation}\label{eq:thaw-scale}
    w_a \approx -1.58(1 + w_0)~.
\end{equation}  
This `calibrated thawing' relation provides a form that acts as a good approximation for the thawing dynamics. However, it also allows the equation of state to cross the phantom line ($ w = -1 $), which is unphysical for quintessence models \cite{Caldwell:1999ew,Cline:2003gs,Vikman:2004dc}.
This occurs because \cref{eq:thaw-scale} is designed to approximate the expansion rate $H(z)$ and distance measures $D(z)$ at sub-percent precision~\textemdash~precisely the quantities probed by cosmological observations~\textemdash~but does not necessarily approximate $w(z)$ itself \cite{dePutter2008}.  

Nevertheless, it is possible to describe thawing dynamics while ensuring that  $w > -1$ at all times. Following \cite{Linder:2007wa,Linder:2015zxa} (see also \cite{Crittenden:2007yy}), the evolution of the thawing equation of state can be parameterized by the algebraic expression  
\begin{equation}\label{eq:thaw-algebraic}
    1 + w(a) = (1 + w_0)\, a^p \left(\frac{1 + b}{1 + b a^{-3}}\right)^{1 - p/3}~,
\end{equation}  
where  $p$ and $w_0$ are free parameters, and $ b = 0.5$ \cite{Linder:2015zxa}. Notably, this formulation, referred to as `algebraic thawing' is more general, where the case $p=1$ and \cref{eq:thaw-scale} were found to yield nearly identical late-time constraints as shown in Appendix A of \cite{DESI:2024kob}.

\begin{figure}
    \centering
    \includegraphics[width=1\linewidth]{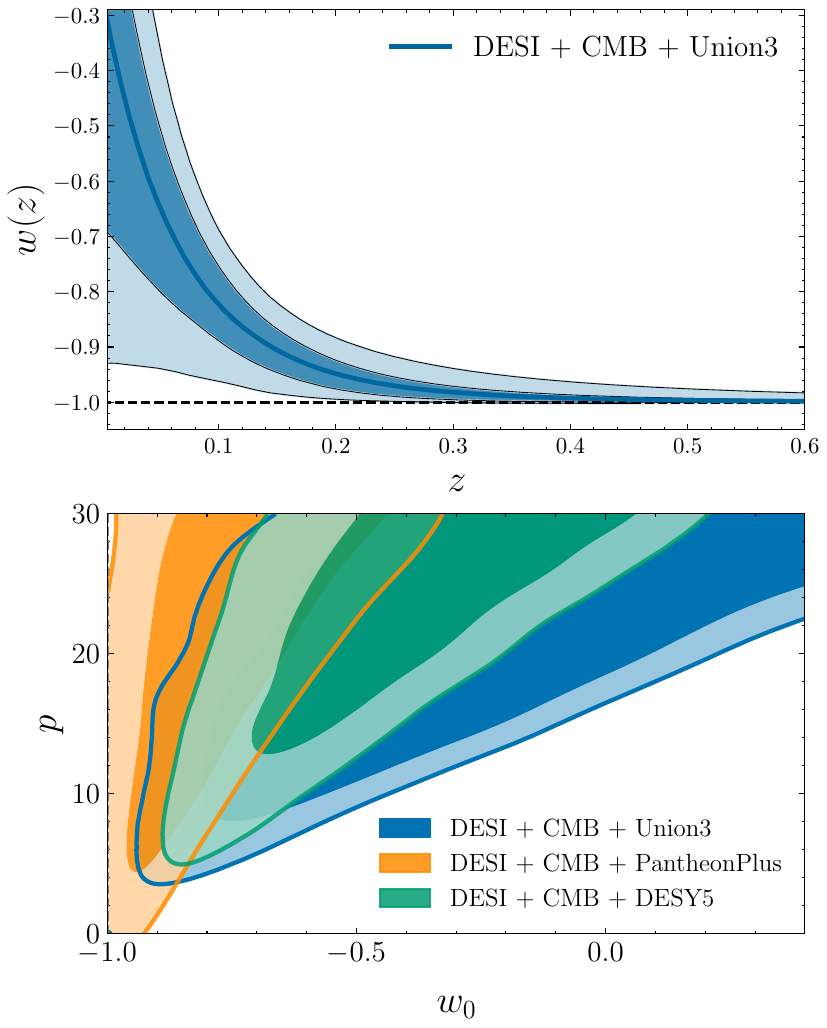}
    \caption{Constraints on the algebraic thawing functional form, as described in \cref{eq:thaw-algebraic}, that restricts $w(z) \geq -1$. The top panel shows the reconstructed evolution of $w(z)$ with shaded regions representing the confidence contours at $68\%$ and $95\%$, demonstrating the deviation from $\Lambda$CDM at low redshifts. The bottom panel illustrates the posterior distributions of $w_0$ and $p$ for different SNe combinations.}
    \label{fig:alge_thaw}
\end{figure}

The reconstructed posterior distribution of $w(z)$ for the thawing class is shown in the top panel of \cref{fig:alge_thaw} for the DESI+CMB+Union3 data combination. This assumes the algebraic form, which enforces $w\geq-1$, and where we have marginalized over the parameter $p$. However, mild degeneracies with $w_0$ leave the posteriors for $p$ largely unconstrained. In particular, it is seen that large values of $p$ are allowed by the data, resulting in our posterior hitting the prior bound $p=30$, as shown in the bottom panel of \cref{fig:alge_thaw}. However, we do not extend our analysis to larger values of $p$, as numerical complications can arise when dealing with DE models with very rapidly varying $w(a)$, particularly in the treatment of perturbations and CMB lensing. 

The calibrated thawing relation \cref{eq:thaw-scale} yields no significant improvement in fit, as seen from the $\Delta\chi^2$ values in \cref{tab:best-fit2} and from the posteriors in \cref{fig:physics-focused} being consistent with $w_0=-1$, except for DESY5. The algebraic thawing parametrization in \cref{eq:thaw-algebraic} can improve the fit with respect to the standard model (\lcdm), achieving a $\Delta\chi^2=-2.9$ for the DESI+CMB+Union3 data combination. Substituting the PantheonPlus data with Union3 or DESY5 yields a $\Delta\chi^2=-6.9$ and $-13.2$, respectively. However, this improvement in fit comes at the cost of including two additional degrees of freedom. 

To illustrate that \cref{eq:thaw-algebraic} correctly captures the phenomenology of thawing fields, and better quantify how the constraints would translate into constraints on the physical parameters of the theory, we consider one physically motivated, axion-like potential \cite{Frieman:1995pm,Kaloper:2005aj,Marsh:2015xka}
\begin{equation}
V(\varphi) = m^2_a f^2_a \left[ 1 + \cos(\varphi/f_a) \right] \, , 
    \label{eq:axion}
\end{equation}  
where $m_a$ denotes the mass of the boson particles related to the scalar field, and $f_a$ is regarded as the effective energy.  Depending on the initial conditions, the axion cosine potential exhibits two distinct behaviors: the standard quadratic regime, the effective mass is positive ($m_{\rm eff}^2 > 0$), and the potential can be approximated by a quadratic form near its minimum, where the effective mass is defined as $m_{\rm eff}^2 = \frac{d^2 V}{d\varphi^2}$.  Whereas, the hilltop regime \cite{Dutta:2008qn} is, characterized by a negative effective mass ($m_{\rm eff}^2 < 0$), when the field begins its evolution near the maximum of the potential (i.e., at $\varphi = 0$) and rolls down toward the minimum at $\varphi = \pi f_a$. We refer the reader to \cref{sec:quintessence} for more details on the model and its implementation in \texttt{class}.

\begin{figure}
    \centering
    \includegraphics[width=1.0\linewidth]{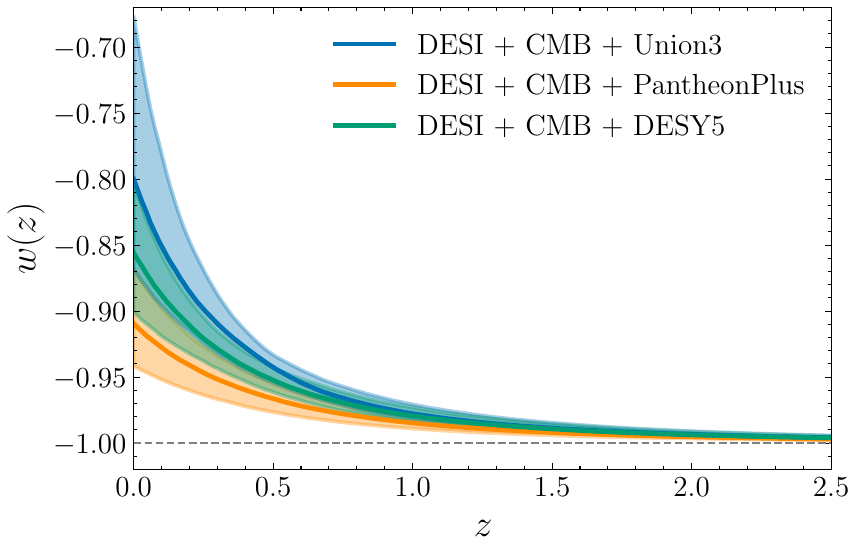}
    \caption{Marginalized constraints on the equation of state parameter, $w(z)=P/\rho$, for the axion-like potential given by \cref{eq:axion}}
    \label{fig:ula-1d-plot}
\end{figure}

In \cref{fig:ula-1d-plot}, we report the marginalized posterior distribution for the equation of state parameter associated with the scalar field potential in \cref{eq:axion}, obtained using DESI, CMB, and three SNe compilations and obtain the following constraints for the physical mass: $\log_{10} (m_a c^2/\mathrm{eV}) = -32.67^{+0.23}_{-0.25}$ (PantheonPlus), $-32.50^{+0.28}_{-0.30}$ (Union3) and $-32.63^{+0.26}_{-0.30}$ (DESY5) and effective energy scale: $\log_{10} (f_a/M_\mathrm{Pl}) = -0.13^{+0.33}_{-0.29}$ (PantheonPlus), $-0.29^{+0.63}_{-0.35}$ (Union3) and $-0.09^{+0.66}_{-0.40}$ (DESY5). 
The constraints indicate that the field starts in the hilltop regime, with initial conditions of $\varphi_i/f_a \sim 0.7-1.0$, rolls down the potential, and reaches the present value of $\varphi_0/f_a \sim 1.1–1.4$, traversing approximately $\Delta \varphi \sim (0.2-0.4) M_{\rm Pl}$.

\subsection{Emergent dark energy}
The second family of DE models that we consider is the \textit{emergent} class, where dark energy had a vanishing presence during most of cosmic history, and only `emerges' in recent times. Following \cite{Li:2019yem,Li:2020ybr}, we parametrize the equation of state as
\begin{equation}
    w(z)=-1-\frac{\Delta}{3\ln{(10)}}\left[1+\tanh\left(\Delta\log_{10}\left(\frac{1+z}{1+z_t}\right)\right)\right]~. \label{eq:GEDE}
\end{equation}
The parameter $\Delta$  determines the steepness of the transition in $w(z)$ and the transtion redshift parameter $z_t$ is determined by the equality $\rho_{\rm DE}(z_t)=\rho_m(z_t)$.
The phenomenology we are trying to capture is that of abrupt 
changes in the equation of state, $w(z)$, driven by physical mechanisms, such as second-order phase transitions \cite{Parker:2000pr,Caldwell:2005xb,Banihashemi:2020wtb}.

\begin{figure}
    \centering
    \includegraphics[width=0.99\linewidth]{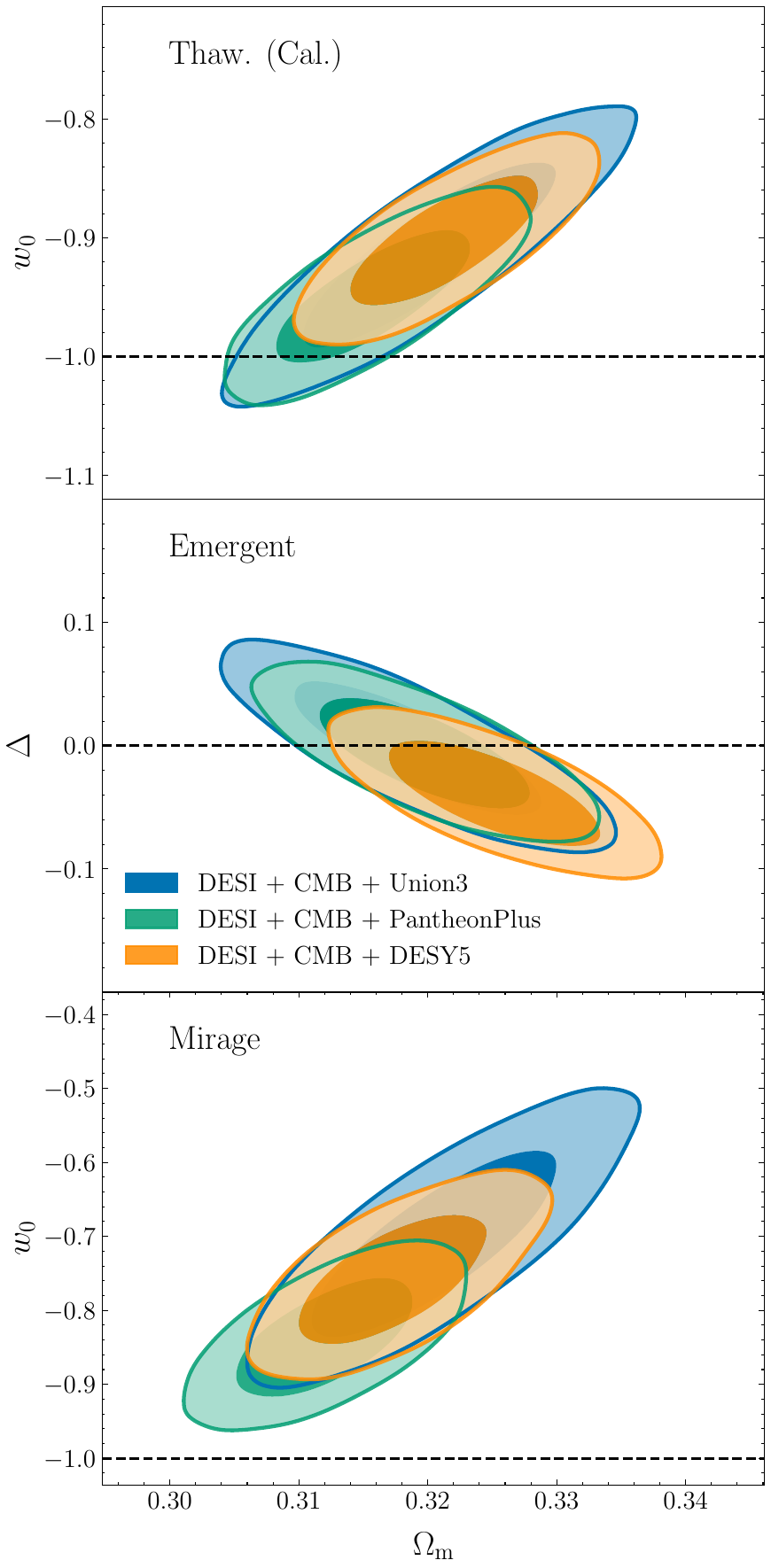}
    \caption{Constraints on three dark energy model classes: calibrated thawing, emergent, and mirage, each with one additional degree of freedom compared to $\Lambda$CDM. The contours represent 68$\%$ and 95$\%$ confidence regions for different SNe combinations: PantheonPlus (blue), Union3 (orange), and DESY5 (green).}
    \label{fig:physics-focused}
\end{figure}

Despite hints of the sharp emergence of dark energy in recent times from non-parametric reconstructions, the DESI+CMB+SNe constraints on $\Delta$, as shown in the middle panel of \cref{fig:physics-focused},  indicate that such an emergent behavior is not statistically favored over $\Lambda$CDM, given the assumed $w(z)$. Note that while \cref{eq:GEDE} can mimic the emergence of dark energy, it is limited by its inability to cross $w = -1$ or, equivalently, introduce a bump in $\fde(z)$; a feature that seems to be favored by the data. 
In principle, one can formulate an emergent dark energy model characterized by an effective equation of state that can cross  $w(z)=-1$. Such behavior may be realized through the coupling of emergent dark energy with the dark matter sector \cite{Turner:1984ff,Barnes:2005bn,Zimdahl:2005bk}.

\subsection{Mirage dark energy} \label{sec:mirage}

The last and more phenomenological class of models which we consider is that of \textit{mirage} dark energy \cite{Linder:2007ka}. This refers to models in the $w_0w_a$ plane (see \cref{fig:w0wa-plane}) approximately living along the line
\begin{equation}\label{eq:mirage}
    w_a\approx -3.66(1+w_0)\ .
\end{equation}
 The mirage class is designed to describe a subset of dynamical dark energy models that preserve the distance to the surface of the last scattering as predicted by \lcdm, a parameter tightly constrained by the CMB \cite{Linder:2007ka}. The name `mirage' stems from the fact that these models would mimic $\Lambda$, yielding $\langle w\rangle \sim -1$ when fitting a constant $w$ to observations, as it could be seen in Table V \cite{DESI.DR2.BAO.cosmo} in DR2 and Table III in \cite{DESI2024.VI.KP7A} for DR1 comparison. The mirage direction fully captures the DE phenomenology suggested by the data, with merely one degree of freedom $w_0$ that quantifies the strength of the mirage, with $w_0 = -1 $ corresponds to $\Lambda$CDM where the mirage is real.  This mirage effect is also expected to persist in the growth of cosmic structures, provided that general relativity remains unmodified \cite{Linder:2005hc,Francis:2007qa,Linder:2007ka}. As noted in \cite{DESI:2024kob}, by reducing the late-time dark energy density (i.e., increasing $\Omega_{\rm m}$), one can make $w_0$ even less negative ~\textemdash~ and correspondingly $w_a$ more negative ~\textemdash~  enhancing the mirage effect. For comparison, DESI+CMB+Union3 prefers $\Omega_{\rm m} \approx 0.304$ in \lcdm\, which increases to $\Omega_{\rm m} \approx 0.327$ in $w_0w_a$ which essentially lies along the mirage direction. 
From the data viewpoint, the mirage line in \cref{fig:w0wa-plane} can be seen as the `principal component', or `axis' in the $w_0w_a$ plane carrying the most meaningful information, i.e., the eigenvector with the highest eigenvalue. Despite effectively reducing the dimensionality of the DE phenomenology, the exact physical mechanism for such rapid emergence of dark energy ($w(a)\ll-1$ )  remains unclear (see \cite{Linder:2024rdj} for more discussion).

\subsection{Model Comparison}
\label{sec:model-comp}
In \cref{fig:physics-focused}, we show the constraints on the single additional parameters of the three DE classes. In the calibrated thawing (top panel), there is a mild deviation observed with the DESY5 dataset, but the other two SNe~Ia datasets indicate overall consistency with $\Lambda$CDM. The emergent (middle panel) exhibits a similar trend, with constraints on $\Delta$ reflecting no significant departures from the standard cosmological framework ($\Delta=0$). In contrast, the mirage class (bottom panel) demonstrates deviations from the value of $w_0 = -1$ across all three SNe~Ia datasets.

We also make a quantitative comparison, examining the deviance information criterion (DIC)\citep{spiegelhalter2002bayesian,liddle2007information}, as defined below, along with the $\dchisq$. The former complements the $\dchisq$ by accounting for model complexity, which does not take into account the number of additional degrees of freedom in a particular model and could be made arbitrarily low if sufficient parameters were added.
The DIC is defined as 
\begin{equation}
   { \rm{DIC}}\,\equiv \, D(\Bar{\theta})+2p_{\rm D}\,=\,\overline{D(\theta)}+p_{\rm D},
   \label{eq:DIC}
\end{equation}
where $ p_{\rm D}\,=\,\overline{D(\theta)}-D(\Bar{\theta})$ is a penalty term, and $D(\Bar{\theta})\,=\,-2\, {\rm{ln}}\, \mathcal{L}+C$ is the `deviance' of the likelihood, with constant $C$ vanishing in $p_{\rm D}$. In practice, we use
 \begin{equation}
     p_{\rm D}\,=\,\overline{\chi^2(\theta)}-\chi^2(\overline{\theta})\,,
 \end{equation}
and $p_{\rm D}$ becomes effectively equivalent to the number of extra parameters in the limit of parameters that are well-constrained with respect to their prior.
 We consider the DE classes in this section and the $w_0w_a$ parametrization, for the data combinations DESI+CMB+SNe~Ia. The comparisons are made between each model class and the $\Lambda$CDM model, and the key metric for each comparison being the DIC. The $\Delta{\rm DIC}$ (and  $\dchisq$) values are reported in \cref{tab:best-fit2}, with a preference for the more complex model indicated by negative values, and for the simpler model, in this case $\Lambda$CDM, by positive values. For nested models, a decrease ($\Delta$DIC $< 0$) of at least 2 is required for a not `insignificant' improvement, and up to 5 constitutes a `positive' preference over $\Lambda$CDM. A decrease of up to 10 is considered a `strong' preference, and beyond this, the preference is `decisive'~\cite{Grandis:2016fwl}.
However, for classes that are not nested within each other, there is no absolute scale for comparison and can only be quantitatively compared against $\Lambda$CDM. 

The $w_0w_a$ parametrization achieves $-\dchisq\sim10.7- 21.0$ and $-\Delta{\rm DIC}\sim6.8-17.2$, indicating that it is strongly to decisively preferred over the standard $\Lambda$CDM.
Comparatively, the calibrated thawing performs poorly. The algebraic thawing class improves the fit slightly more, and since the $\Delta{\rm DIC}$ is consistently larger than for calibrated thawing, the improvement in $\chi^2$ must be sufficient to reconcile the former's second additional degree of freedom.
The emergent dark energy class shows no significant improvement in fit, faring even worse than the thawing class.
The final mirage class attains $-\dchisq\sim10.5-20.7$, comparable to the $w_0w_a$ parametrization, as well as $-\Delta{\rm DIC}\sim9.1-18.7$. This is perhaps unsurprising, given how closely the mirage direction aligns the $w_0w_a$CDM constraints.

The advantage of the thawing and emergent classes is their connection to a physical interpretation, which is fairly straightforward, while for the mirage class less so. It is also important to note that the model comparison metrics used in this section serve both to quantify the data’s preference for each model—providing an absolute scale in the case of nested models—and to facilitate a relative ranking among non-nested models, such as the algebraic thawing and $w_0w_a$ parameterizations. 

\begin{table}
    \centering
    \small
    \caption{$\Delta \chisq_\mathrm{MAP} \equiv \chisq_{\rm model}-\chisq_{\Lambda\rm CDM}$ and $\Delta{\rm DIC}\equiv {\rm DIC}_{\rm model}-{\rm DIC}_{\rm \Lambda CDM}$ values for various data combinations, including DESI, CMB with each different SNe~Ia, and
    DE classes, namely: Thawing (Calibrated and Algebraic: `Thaw.~(Cal.)' and `Thaw.~(Alg.)', respectively), Emergent and Mirage. The minimum $\chisq$ values were obtained using the \texttt{iminuit} \cite{iminuit} minimizer.}
    \begin{ruledtabular}
    \begin{tabular}{l l l l l l l} 
        DESI+CMB: &\multicolumn{2}{c}{{+PantheonPlus}}&\multicolumn{2}{c}{{+Union3}} &\multicolumn{2}{c}{+DESY5}\\ 
        \hline
        \textbf{DE classes} & \multicolumn{6}{c}{$\Delta {\rm DIC}$ ($\Delta \chi^2$)} \\ \hline
        
        \hline
        Thaw. (Cal.)     & $+0.4$  &$(-1.6)$ & $-0.6$  &$(-2.5)$ & $-5.8$  &$(-7.1)$ \\  
        Thaw. (Alg.) & $-1.0$  &$(-2.9)$& $-4.6$  &$(-6.9)$  & $-10.1$ &$(-13.2)$\\  
        
        Emergent       & $+2.1$  &$(-0.05)$   & $+1.8$ &$(-0.1)$    & $+0.2$  &$(-1.5)$\\  
        Mirage     & $-9.1$ &$(-10.5)$ & $-13.8$ &$(-16.2)$ & $-18.7$ &$(-20.7)$\\  
        \hline
        ${w_0w_a}$ & $-6.8$ &$(-10.7)$ & $-13.5$ &$(-17.4)$ & $-17.2$ &$(-21.0)$\\ 
    \end{tabular}
    \end{ruledtabular}
    
    \label{tab:best-fit2}
\end{table}

\subsection{Is there evidence for phantom crossing?}
\label{sec:phantom}

Both the parametric and non-parametric methods discussed in \cref{sec:params} and \cref{sec:non-params} indicate a possible crossing of the phantom divide line; however, as illustrated in \cref{fig:xtreme_mock}, this does not guarantee that the crossing is genuine. Specifically, the apparent crossing in the $w_0w_a$ parameterization may be spurious to match observables, raising the question of whether $w(a)$ truly crosses $-1$ or if the behavior is simply an artifact of the parameterization. 

To address this, we analyze the behavior of thawing quintessence using the algebraic model described by \cref{eq:thaw-algebraic}, which enforces $w(z) \ge -1$.  Although the algebraic thawing model, restricted by our prior $p < 30$, yields a better fit to the data than $\Lambda$CDM by $3 \lesssim -\Delta\chi^2 \lesssim 13$ and $-1 < -\Delta\rm{DIC} < 10$, it is considerably less favored than the $w_0w_a$ model. This overall preference for $w_0w_a$CDM over algebraic-thawing model, however, cannot be straightforwardly converted into p-values or n-sigma levels because algebraic thawing and $w_0w_a$CDM models are not nested.

To achieve $-\Delta\chi^2$ comparable to $w_0w_a$ with thawing models would require an exceptionally fine-tuned potential $V(\varphi)$ \cite{Wolf:2024eph}, precise initial field settings \cite{Shlivko:2024llw}, or `data-informed' prior choices \cite{Payeur:2024dnq}, resulting in $w(z \gtrsim 0.3) = -1$ followed by a rapid increase to $w(z \lesssim 0.3) > -1$.
This suggests that a sharp increase followed by a decrease in dark energy density may be a necessary feature, since models that do not cross the phantom divide tend to underperform compared to those that do. The specific behavior of $w(z)$ suggested by the data~\textemdash~a phantom crossing from $w(z\gtrsim 0.5)<-1$ to $w(z\lesssim 0.5)>-1$~\textemdash~ is not predicted by any of the simplest and most studied extensions of \lcdm\ (see also the recent discussion in \cite{Linder:2024rdj}).

While the phantom crossing can be seen as theoretically challenging due to stability issues, for example, within the framework of minimally-coupled scalar fields, obtaining such behavior is not difficult in extended theoretical frameworks. For example, phantom crossing can arise in models where dark energy possesses multiple internal degrees of freedom, such as multi-field scenarios \cite{Hu:2004kh,Guo:2004fq,Wei:2005nw,Caldwell:2005ai,Cai:2007zv,Cai:2007qw}, non-standard vacuum models \cite{Parker:2000pr,Caldwell:2005xb}, frameworks where dark energy interacts with dark matter \cite{Amendola:1999er,Billyard:2000bh,Amendola:2003wa,Nojiri:2005sx,Clemson:2011an,Shafieloo_2017}, and modified theories of gravity \cite{Carvalho:2004ty,Hu:2007nk,PhysRevLett.96.061303,Anisimov:2005ne,Nesseris:2006jc,Deffayet:2010qz,Pujolas:2011he,Ye:2024ywg,Wolf:2024stt,Yang:2024kdo,Christiansen:2024hcc}. 
Because of the aforementioned multiple internal degrees of freedom in these models, the effective, observable equation of state $w(z)$ can cross the phantom divide even though the null-energy condition is not violated.  Whether a compelling theoretical mechanism \textemdash~ one that does not require many extra degrees of freedom or exotic assumptions \textemdash~ can be constructed to cross the phantom divide in the way suggested by data remains an open question, although some models have recently been put forward as viable in this regard \cite{Wolf:2024stt,Ye:2024ywg}.

\section{Conclusions}\label{sec:conclusions}
This work presents constraints on dark energy from DR2 BAO, in combination with cosmic microwave background, and type Ia supernova data. We began our study by summarizing and expanding upon the $w_0w_a$CDM analysis presented in \cite{DESI.DR2.BAO.cosmo}. Under the assumption of a linearly evolving $w(a)=w_0+w_a(1-a)$, the latest DESI+CMB results indicate a $\simeq3\sigma$ deviation from \lcdm. The data shows a clear preference for the ($w_0 > -1$, $w_a < 0$) quadrant, and in particular $w_0+w_a<-1$, implying a past phantom-like equation of state transitioning to $w(z) > -1$ today. The reconstructed $Om(z)$ and deceleration parameter $q(z)$ also show clear deviations from $\Lambda$CDM, reinforcing the case for evolving dark energy.

To assess the robustness of our findings, we conducted a series of analyses: i) varying the redshift dependence of $w(z)$, by considering various parametrizations (\cref{sec:alt-par}), and ii) studying the improvement in fit as more freedom is given to the dark energy characteristics (\cref{sec:crossing}). As in DR1, the results are rather stable under changes in the assumed form for $w(z)$, and the data does not seem to require more degrees of freedom in $w(z)$, beyond $w_0w_a$, as shown in \cref{fig:delta_chi2_N_crossing} (see also \cite{DESI:2024aqx}). 

Next, we implemented two non-parametric reconstruction techniques and applied them to the redshift-dependent equation of state $w(z)$ and dark energy density $\fde(z)$, in order to allow more flexibility than that available in the parametric methods.
Overall, the constraints support the evolution indicated by the $w_0w_a$ parametrization, giving the tightest constraints at low redshifts, where they display a preference for a deviation with $w(z)>-1$, while suggesting a crossover to the phantom regime at higher redshift. The low redshift deviation is evidently independent of the chosen binning variable, although the constraints remain within $2\sigma$ of $\Lambda$CDM at higher redshifts. Gaussian process regression is better able to localize the redshift where the crossing should occur, around $z\sim0.5$.

In order to provide possible interpretations for the physical origin of the observed deviation, three model classes were considered, each endowed with a different dynamical behavior and motivated to various degrees by physical theory. 
The thawing and emergent classes are the least well-supported, indicating that the data might not favor dark energy evolution that arises from, respectively, either minimally-coupled scalar field models or emergent behavior in energy density.
In contrast, the mirage class performs remarkably well, capturing DE phenomenology with just one additional degree of freedom, which warrants an inquiry into whether any underlying physics or systematic effects could explain this mirage.

In summary, irrespective of the parametric/non-parametric methods used, the evidence of deviation from $\Lambda$CDM is significant. Our findings suggest that the canonical $w_0w_a$ parametrization effectively captures the essence of dark energy evolution in our study.
Decisive tests of dark energy and its possible deviations from the $\Lambda$CDM model will require a combination of complementary probes.  The forthcoming DESI data releases, including constraints from redshift space distortions and peculiar velocities, will offer crucial insights into the nature of dark energy and gravity. The upcoming SNe measurements from the ZTF survey~\cite{ZTFSNDR2,ZTFoverview},  the Vera C. Rubin Observatory ~\cite{LSSTWFD, LSSTDD} and Nancy Grace Roman Space
Telescope ~\cite{Roman_report} will extend the Hubble diagram probed by DESI to very low redshifts, improving constraints on $w_0$. Meanwhile, data from Euclid~\cite{Euclid_report} and Rubin will serve as an important cross-check of DESI’s findings, helping to assess the impact of potential systematics. Finally, next-generation CMB experiments will further tighten constraints on early-universe parameters, breaking degeneracies with late-time observables. 
With these advancements, the next decade promises to determine if we are entering a new era in modern cosmology that necessitates a paradigm shift.

\section{Data Availability}
The data used in this analysis will be made public along the Data Release 2 (details in \url{https://data.desi.lbl.gov/doc/releases/}).
The data points corresponding to the figures from this paper will be available
in a Zenodo repository.

\acknowledgments

The authors thank Eric Linder for his valuable discussions and Robert Crittenden and Kazuya Koyama for their detailed comments. R.C. is funded by the Czech Ministry of Education, Youth and Sports (MEYS) and European Structural and Investment Funds (ESIF) under project number CZ.02.01.01/00/22\_008/0004632. A.S. would like to acknowledge the support by National Research Foundation of Korea 2021M3F7A1082056. 
MI acknowledges that this material is based upon work supported in part by the Department of Energy, Office of Science, under Award Number DE-SC0022184, and also in part by the U.S. National Science Foundation under grant AST2327245.
CGQ acknowledges support provided by NASA through the NASA Hubble Fellowship grant HST-HF2-51554.001-A awarded by the Space Telescope Science Institute, which is operated by the Association of Universities for Research in Astronomy, Inc., for NASA, under contract NAS5-26555.

This material is based upon work supported by the U.S.\ Department of Energy (DOE), Office of Science, Office of High-Energy Physics, under Contract No.\ DE–AC02–05CH11231, and by the National Energy Research Scientific Computing Center, a DOE Office of Science User Facility under the same contract. Additional support for DESI was provided by the U.S. National Science Foundation (NSF), Division of Astronomical Sciences under Contract No.\ AST-0950945 to the NSF National Optical-Infrared Astronomy Research Laboratory; the Science and Technology Facilities Council of the United Kingdom; the Gordon and Betty Moore Foundation; the Heising-Simons Foundation; the French Alternative Energies and Atomic Energy Commission (CEA); the National Council of Humanities, Science and Technology of Mexico (CONAHCYT); the Ministry of Science and Innovation of Spain (MICINN), and by the DESI Member Institutions: \url{https://www.desi. lbl.gov/collaborating-institutions}. 

The DESI Legacy Imaging Surveys consist of three individual and complementary projects: the Dark Energy Camera Legacy Survey (DECaLS), the Beijing-Arizona Sky Survey (BASS), and the Mayall z-band Legacy Survey (MzLS). DECaLS, BASS and MzLS together include data obtained, respectively, at the Blanco telescope, Cerro Tololo Inter-American Observatory, NSF NOIRLab; the Bok telescope, Steward Observatory, University of Arizona; and the Mayall telescope, Kitt Peak National Observatory, NOIRLab. NOIRLab is operated by the Association of Universities for Research in Astronomy (AURA) under a cooperative agreement with the National Science Foundation. Pipeline processing and analyses of the data were supported by NOIRLab and the Lawrence Berkeley National Laboratory. Legacy Surveys also uses data products from the Near-Earth Object Wide-field Infrared Survey Explorer (NEOWISE), a project of the Jet Propulsion Laboratory/California Institute of Technology, funded by the National Aeronautics and Space Administration. Legacy Surveys was supported by: the Director, Office of Science, Office of High Energy Physics of the U.S. Department of Energy; the National Energy Research Scientific Computing Center, a DOE Office of Science User Facility; the U.S. National Science Foundation, Division of Astronomical Sciences; the National Astronomical Observatories of China, the Chinese Academy of Sciences and the Chinese National Natural Science Foundation. LBNL is managed by the Regents of the University of California under contract to the U.S. Department of Energy. The complete acknowledgments can be found at \url{https://www.legacysurvey.org/}.

Any opinions, findings, and conclusions or recommendations expressed in this material are those of the author(s) and do not necessarily reflect the views of the U.S.\ National Science Foundation, the U.S.\ Department of Energy, or any of the listed funding agencies.

The authors are honored to be permitted to conduct scientific research on Iolkam Du’ag (Kitt Peak), a mountain with particular significance to the Tohono O’odham Nation.



\bibliographystyle{mod-apsrev4-2}
\bibliography{references, DESI_supporting_papers}


\appendix

\section{Bayesian Model Comparison}\label{sec:Model-Comparison}
Here we discuss the Bayesian model comparison and compute Bayes factor between the $w_0w_a$ and $\Lambda$ models, which is given by the corresponding ratio of their evidence $\mathcal{Z}$ under a given data set: $B_{w_0w_a \Lambda} = \mathcal{Z}_{w0w_a}/\mathcal{Z}_\Lambda$. 
In practice, we compute Bayes factors using the nested sampler \texttt{polychord} \cite{Handley:2015fda}, employing the same priors\footnote{We substitute $\theta_{\rm MC}$ with $H_0$ to avoid numerical issues arising from the shooting method.} and the Boltzmann solver (CAMB) as in the posterior analysis discussed in the main text. The evidence is then estimated using the \texttt{anesthetic} package \cite{Handley:2019mfs}.

The logarithm of the evidence ($\log \mathcal{Z}$) can be expressed as the contribution from two terms in the form \cite{Hergt:2021qlh}
\begin{equation}
    \log \mathcal{Z} = \langle \ln \mathcal{L} \rangle_\mathcal{P} - \langle \ln (\mathcal{P} /\pi) \rangle_\mathcal{P} 
\end{equation}
where, $\mathcal{L}$ is the likelihood, $\mathcal{P}$ is the posterior, $\pi$ represents the prior distribution and $\langle \rangle_\mathcal{P} $ is the posterior weighted average. The posterior average of the log-likelihood $\langle\ln\mathcal{L}\rangle_\mathcal{P}$, removes the prior-dependent Occam`s penalty, $\langle \ln (\mathcal{P} /\pi) \rangle_\mathcal{P}$, contribution from the log-evidence to provide a quantitative assessment of how well the model fits the data and can be considered the Bayesian equivalent of $\chi^2$. 

In \cref{tab:bayes}, we report the differences between the $w_0w_a$CDM  and $\Lambda$CDM models, for values of $\ln B$ ($=\Delta \ln \mathcal{Z})$, $\Delta \langle \ln \mathcal{L} \rangle_\mathcal{P}$, and $\Delta \langle \ln (\mathcal{P} /\pi) \rangle_\mathcal{P}$ for each data combination. The Bayesian evidence ratio indicates that the support for the $w_0w_a$ model increases with the latest DR2 release. On Jeffreys’ scale \cite{Jeffreys:1939xee, Trotta:2005ar}, the DESY5 combination indicates almost a strong preference ($\leq5$) for the $w_0w_a$CDM model when combined with DR2 compared to a moderate preference ($\leq 2.5$) with the DR1 dataset, while Union3 shows a moderate preference for $w_0w_a$ compared to the weak preference observed when combined with the DR1 dataset. In contrast, PantheonPlus provides inconclusive evidence, showing a preference for $\Lambda$, though this preference diminishes in the change from DR1 to DR2.  In all three cases, we observe that the trends in $\Delta \langle\ln\mathcal{L}\rangle_\mathcal{P}$ remain consistent with the frequentist $\Delta \chi^2$ results presented in \cref{tab:best-fit2}, with a clear improvement in the likelihood of the fit when switching from the DR1 to the DR2 dataset. We note that since the Bayesian evidence value includes likelihood contributions across a range of possible parameter values, it is less susceptible to random fluctuations in the data leading to a better fit at a specific point in parameter space by chance. Therefore, $\ln B$ or $\Delta \langle \ln \mathcal{L} \rangle_\mathcal{P}$ comparisons should be less noisy and more robust than $\Delta \chi^2$.

A larger Occam's penalty (i.e., a greater compression from prior to posterior) is typically expected with more informative data. However, this is not the case for the DESY5 dataset, although it is a better fit for the data. Even so, all of the Occam penalty factors for DESI + CMB + SNe~Ia data are roughly consistent with each other, given the estimated error on these values.

Please note that in the \cite{DESI2024.VI.KP7A} used the Boltzmann solver \texttt{class} with a different set of priors and CMB likelihood. For a better comparison, we recomputed evidence for DESI DR1 BAO using consistent priors and methodology as described in the \cref{sec:data}.

We remind readers that Bayesian model comparison relies on the prior, this is especially important in the case of testing phenomenological models like $w_0w_a$CDM, as the prior chosen for the extra parameters is only phenomenologically justified. Occam's penalty can be adjusted by choosing different priors. Generally, a wider prior tends to favor the simpler model,  in this case is the $\Lambda$CDM. However, the priors chosen in Tab.~\ref{tab:priors} are broad enough for $w_0w_a$, and the posterior of these parameters is well constrained inside this prior when the datasets are combined. A different choice of prior on these parameters could change the ranking, such that $\Lambda$CDM would be favored over $w_0w_a$CDM for all data combinations, but to do this in the most extreme case (DESI DR2 BAO + CMB + DESY5) the prior range would need to be expanded by a factor of more than 10 times for both $w$ parameters, leading to a greater than 100 times expansion in prior volume. Such a prior would not doubt be considered unphysical and unreasonable by most.

\begin{table}
\caption{\label{tab:bayes}Bayesian evidence $\ln B_{w_0 w_a \Lambda}$, posterior average of log-likelihood $\langle \ln \mathcal{L} \rangle_{\mathcal{P}}$, and Occam’s penalty $\langle \ln (\mathcal{P} / \pi) \rangle_{\mathcal{P}}$ for different supernova datasets (PantheonPlus, Union3, and DESY5) in combination with CMB and DESI BAO measurements. We report results for DESI DR2 BAO + CMB  and DESI DR1 BAO + CMB with associated uncertainties for comparison.  All reported values correspond to the difference between the $w_0w_a$CDM and the $\Lambda$CDM.}
\begin{ruledtabular}
\begin{tabular}{lccc}
 & +PantheonPlus & +Union3 & +DESY5 \\
\hline
\multicolumn{2}{c}{\textbf{DESI DR2 BAO + CMB}} \\
\hline
$\ln B_{w_0w_a \Lambda}$ & $-0.66 \pm 0.44$ & $3.03 \pm 0.44$ & $4.88 \pm 0.44$ \\
$\Delta \langle \ln \mathcal{L} \rangle_\mathcal{P} $ & $4.53 \pm 0.17$ & $7.78 \pm 0.16$ & $9.42 \pm 0.16$ \\
$\Delta \langle \ln (\mathcal{P} /\pi) \rangle_\mathcal{P} $ & $5.18 \pm 0.43$ & $4.75 \pm 0.43$ & $4.54 \pm 0.44$ \\
\hline
\multicolumn{2}{c}{\textbf{DESI DR1 BAO + CMB}} \\
\hline
$\ln B_{w_0w_a \Lambda}$ & $-1.54 \pm 0.44$ & $1.62 \pm 0.44$ & $2.35 \pm 0.44$ \\
$\Delta \langle \ln \mathcal{L} \rangle_\mathcal{P}$ & $3.34 \pm 0.17$ & $6.03 \pm 0.16$ & $7.67 \pm 0.16$ \\
$\Delta \langle \ln (\mathcal{P} /\pi) \rangle_\mathcal{P} $ & $4.87 \pm 0.43$ & $4.41 \pm 0.43$ & $5.32 \pm 0.43$ \\
\end{tabular}
\end{ruledtabular}
\end{table}

\section{Details of Binning PCA}
\label{sec:BinApp}

Principal component analysis (PCA) is a commonly-used strategy that leverages the number of bins, or additional free parameters, against the uncertainties in the constrained amplitudes, in an attempt to analyse how most efficiently to segment the data~\cite{Huterer_2003,Huterer_2005,Crittenden:2005wj,dePutter2008,Zhao:2017cud}, though the interpretation is not necessarily straightforward, and it remains subject to various caveats~\cite{dePutter2008}.

In this appendix, we lay out the full mathematical expressions for the PCA performed on the binning results. We start by writing the equation of state in terms of the binning amplitude parameters and an initial basis
\be
w(z) =  \sum_{j=0}^N w_j{\mathbf e}_j
\ee
where $w_i$ are the bin amplitude parameters, defined as previously, and ${\mathbf e}_j$ is simply the tanh smoothed top-hat function
\begin{align}
{\mathbf e}_j &\equiv {\mathbf e}(z,z_j,z_{j-1}) \notag \\
&= \frac{1}{2}\left[\tanh\left(\frac{z-z_j}{s}\right)- \tanh\left(\frac{z-z_{j-1}}{s}\right)\right]\,,
\end{align}
which has value one in the interval $z_{j-1} \lesssim z \lesssim z_j$ and zero elsewhere, allowing us to recover the expression in \cref{eq:binning}.

If we diagonalize the covariance matrix of these coefficients to obtain
\be
C^{-1} = O^T\Lambda O\,,
\ee
then the orthogonal matrix $O$ contains the eigenvectors of $C^{-1}$ and the diagonal matrix $\Lambda$ contains the corresponding eigenvalues~\cite{Huterer_2003,dePutter2008}. For convenience, we normalize $O$ to have a determinant of one.

To obtain the new coefficients, we use the rows of $O$ as weights on the originals
\be
q_i = \sum_{j=0}^N O_{ij}w_j\,.
\ee
Since $O$ is orthogonal, i.e. $O^{-1} = O^{T}$, the coefficients and basis will transform in the same way. As such, the new basis functions (see~\cref{fig:PCA_bin}) may be obtained in a similar fashion using~\cite{Huterer_2003,dePutter2008}
\be
{\mathbf e}^\prime_i =  \sum_{j=0}^N O_{ij}{\mathbf e}_j\,.
\ee
The uncertainty in the new parameters is given by the inverse of the eigenvalues~\cite{Huterer_2003}
\be
\sigma_i \equiv \sigma[q_i] = \Lambda^{-1}_{ii}\,.
\ee
While the PCA is no doubt a useful approach, it must be noted that, in general, eigenvectors are formally not well-defined for the inverse covariance matrix, and that the set of eigenvectors found will themselves depend on the binning parametrization and variable. Finally, there is no clear \textit{a priori} interpretation of the size of $\sigma[q_i]$ without making further assumptions about the form of the equation of state~\cite{dePutter2008}.

\section{Details on Gaussian Process Regression}\label{sec:GP-app}

In this appendix, we discuss the details of the methodology and implementation of the Gaussian process regression results presented in \cref{sec:gp}. We adopt a flexible and non-parametric approach to model the dark energy equation of state $w(z)$ utilizing a Gaussian Process with a squared exponential kernel given by

 \begin{equation}\label{eq:w-GP}
    w(z)\sim\mathcal{GP}\left(m(z)=-1,K=k(\sigma_f,\ell_f) \right)~,
\end{equation} 
\begin{equation}\label{Kernel}
     k(x_i,x_j;\sigma_f,\ell_f)=\sigma_f^2~\exp\left(-\frac{(x_i-x_j)^2}{2\ell_f^2}\right)~,
\end{equation}

where $\sigma_f$ denotes the typical deviations of $ w $ from the mean function, which we consider to be the $\Lambda$CDM value $ w(z) = -1 $. The parameter $ \ell_f $ controls the correlation length of samples. We draw samples from a joint Gaussian distribution across a dense uniform grid ranging from $ z = 0 $ to $ z = 10 $ that smoothly go back to $ w(z) \rightarrow -1 $ at high redshifts to maintain numerical stability. We fix $\sigma_f = 1$ for accommodating a broad spectrum of dark energy behaviors and impose a generalized inverse Gaussian prior on $ \ell_f $:

\begin{equation}
f(x \mid p, b) = \frac{(b)^{p/2}}{2 K_p(\sqrt{b})} x^{p-1} e^{-(x + b/x)/2}, \quad x > 0, \label{eq:ellf_prior}
\end{equation}

where $ K_p(\sqrt{b}) $ denotes the modified Bessel function of the second kind. We select $ p=3 $ and $ b=2 $ to penalize low $ \ell_f $ values, thus preventing excessive freedom in $w(z)$, while also constraining high $ \ell_f $ values to avoid oversampling nearly linear functions. To improve the efficiency of our sampling procedure, we introduce a latent variable that directly samples $ w(z_l) $ at $ z_l $ (we fix $z_l = 0.4$) from a Gaussian prior with width $ \sigma_f $. These sampled values are subsequently employed to generate samples of $ w(z) $ from a conditional distribution, facilitating significantly quicker convergence. We have assessed the impact of varying $ z_l $ (over the range  $z \in [0.2, 1.5]$)  and marginalizing $ \sigma_f $ with a uniform prior ($\pi(\sigma_f) = \mathcal{U}[0,2]$), finding our results and conclusions to be reasonably robust.

\section{Quintessence}
\label{sec:quintessence}

We discuss the constraints on a quintessence model with a scalar field $\varphi$ minimally-coupled to gravity, with action given by (e.g.~\cite{Amendola:2015ksp})
\begin{equation}
    S_\varphi = \int d^4 x \sqrt{-g} \left[ - \frac{1}{2} g^{\mu \nu} \partial_\mu \varphi \partial_\nu \varphi - V(\varphi) \right] \, , \label{eq:action-varphi}
\end{equation}
where $g$ is the determinant of the metric $g_{\mu \nu}$. For our purposes, the scalar field potential has a periodic dependence on the field $\varphi$ in the form,
\begin{equation}
V(\varphi) = m^2_a f^2_a \left[ 1 + \cos(\varphi/f_a) \right] \, , 
\end{equation}
where $m_a$ denotes the mass of the boson particles related to the scalar field, and $f_a$ is regarded as the effective energy scale of the theory. This model, also referred to as PNGB in~\cite{Frieman:1995pm}, has since been a recurrent feature in cosmological analyses~\cite{Shajib:2025tpd,Smer_Barreto_2017,dePutter2008,Abrahamse:2007te,Kawasaki:2001bq,Ng:2000di,Coble:1996te,Frieman:1997xf,Viana:1997mt}.
The Klein-Gordon equation derived from the action in~\cref{eq:action-varphi}, in a standard cosmological setting, is
\begin{equation}
    \Ddot{\varphi} + 3H \Dot{\varphi} - m^2_a f_a \sin(\varphi/f_a) = 0 \, ,
\end{equation}
with $H$ being the Hubble parameter. We then perform the following polar transformation on the field variables $(\varphi,\Dot{\varphi})$ (see~\cite{Cedeno:2017sou,LinaresCedeno:2020dte} and references therein),
\begin{subequations}
\begin{eqnarray}
    \sqrt{\frac{2}{3}} \frac{m_a f_a \cos(\varphi/2f_a)}{M_{Pl} H} &=& \Omega^{1/2}_\varphi \cos (\theta/2) \, , \\  
     \frac{\Dot{\varphi}}{\sqrt{6} M_{Pl} H} &=& \Omega^{1/2}_\varphi \sin (\theta/2) \, . \label{eq:polar-transf}
\end{eqnarray}
\end{subequations}
Here, $M_{Pl}$ is the reduced Planck mass, $\Omega_\varphi$ is the density parameter of the scalar field, and $\theta$ is an angular variable directly related to the scalar field equation of state $w_\varphi$,
\begin{equation}
    w_\varphi = \frac{\Dot{\varphi}^2 - 2m^2_a f^2_a \left[ 1 + \cos(\varphi/f_a) \right]}{\Dot{\varphi}^2 + 2m^2_a f^2_a \left[ 1 + \cos(\varphi/f_a) \right]} = - \cos \theta \, . \label{eq:w-varphi}
\end{equation}
Equation~\eqref{eq:w-varphi} directly shows that, for quintessence models, the equation of state only varies in the range $-1 \leq w_\varphi \leq 1$.

As a result, the Klein-Gordon equation of motion can be rewritten as a dynamical system in the new variables, namely,
\begin{subequations}
\label{eq:polar-motion}
\begin{eqnarray}
    \theta^\prime &=& -3 \sin \theta + \sqrt{y^2 - \alpha \Omega_\varphi (1+\cos \theta)} \, , \\
    y^\prime &=& \frac{3}{2} (1+w_{\rm tot}) y \, , \\
    \Omega^\prime_\varphi &=& 3 (w_{\rm tot} + \cos \theta) \Omega_\varphi \, ,
\end{eqnarray}
\end{subequations}
where a prime denotes a derivative with respect to $N = \ln a$, and $w_{\rm tot}$ is the total equation of state of the matter budget of the universe. The boson mass $m_a$ appears implicitly in the definition of the new dynamical variable $y = 2m_a/H$, which directly measures the ratio of the boson mass to the Hubble parameter. Likewise, the effective energy scale $f_a$ appears implicitly in the new parameter $\alpha = 3/(f_a/M_\mathrm{Pl})^2$ and adopt following prior $m_a \in \mathcal{U}[0.1,10]$ and $\alpha \in \mathcal{U}[10^{-6},100]$ to sample mass and effective scale.

To solve the dynamical system~\cref{eq:polar-motion}, we follow the prescription in~\cite{LinaresCedeno:2020dte}. The initial value of $y$ is determined by choosing a value of the boson mass $m_a$, that is, $y_i = 2 m_a/H_i$, while the initial angular variable is given by $\theta_i = (1/5) y_i$, which is an attractor solution at early times. As we will assume that $m_a \sim H_0 \ll H_i$, the initial value of the equation of state is $w_{\varphi i} \simeq -1$, which means that the field $\varphi$ starts its evolution close to a slow-roll regime. Finally, once a value of $\alpha$ is chosen, the initial value $\Omega_{\varphi i}$ is adjusted using a numerical shooting routine inside the Boltzmann solver \texttt{class}  until the desired value of $\Omega_\varphi$ at the present time is obtained. Using the polar transformations~\eqref{eq:polar-transf}, the physical parameters $(m_a,f_a)$, and the initial values $(H_i,\Omega_{\varphi i})$, one can calculate the initial values of the original field variables $(\varphi_i,\Dot{\varphi}_i)$ corresponding to the attractor solution at early times.

\begin{figure}
    \centering
    \includegraphics[width=1\columnwidth]{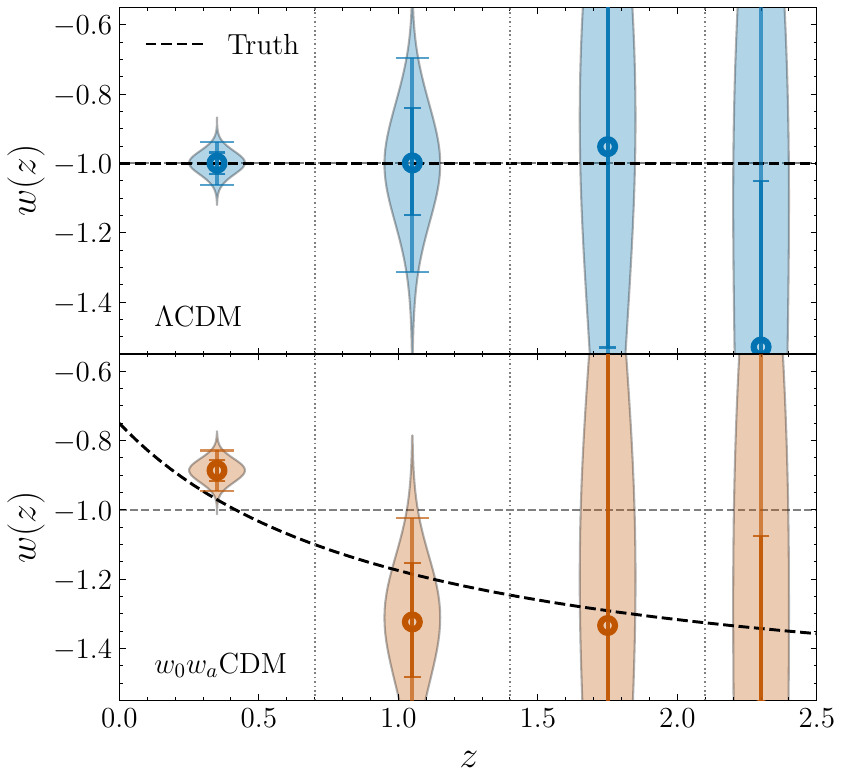}
    \caption{Validation of $w(z)$ binned in 3 uniform bins, for the two mock datasets. The circular data points show the median values of the reconstruction, with 1$\sigma$ and 2$\sigma$ vertical error bars, as well as posterior distributions of the bin amplitude parameters. In both cases, the true evolution of $w$, shown in a black dashed line, is well recovered.} 
    \label{fig:binning-fde-mocks}
\end{figure}

\section{Validation on Mocks}
\label{sec:mock_val}

Non-parametric approaches offer significant advantages by minimizing assumptions about the underlying physical properties of dark energy, allowing for a more flexible and unbiased reconstruction of cosmic expansion and structure growth. 
However, their power comes with the challenge of ensuring robustness, as the lack of an explicit model can introduce degeneracies and reconstruction artifacts. Careful validation using mock datasets is essential to assess the reliability of these techniques
and identify potential biases that may arise due to the methodology itself, ensuring that the inferred constraints are driven by true cosmological signals and not systematic effects.

\begin{figure}
    \centering
    \includegraphics[width=1\columnwidth]{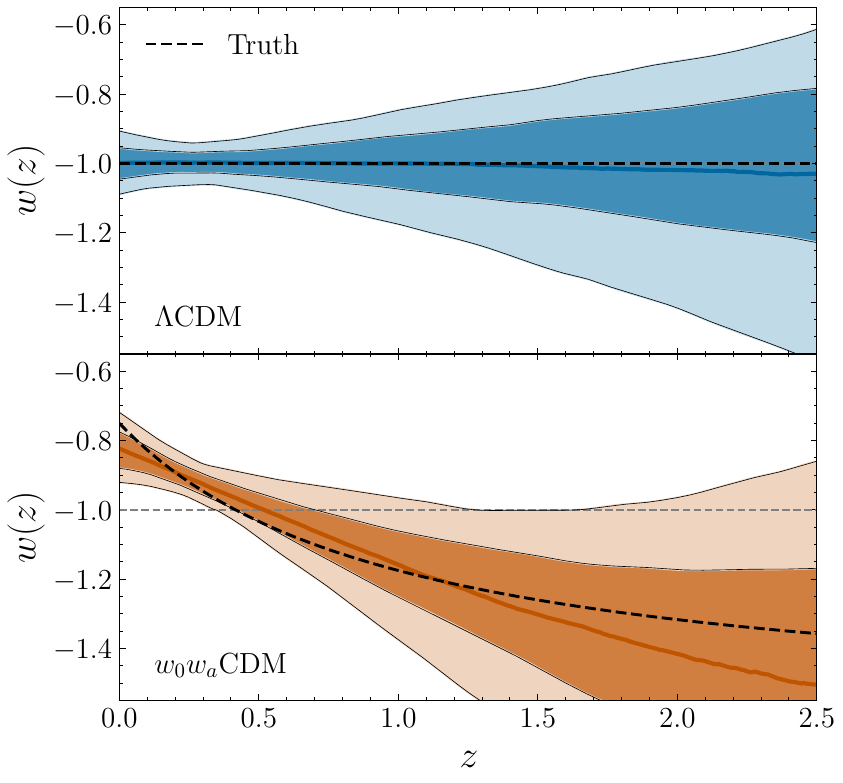}
    \caption{Validation of $w(z)$ reconstructed using GP, for two mock datasets. In both of the mocks, the true $w(z)$ function, used to generate them and shown in a black dashed line, is recovered well within the $1\sigma$ contour.}
    \label{fig:w_GP_mocks}
\end{figure}

\begin{figure}[h!]
    \centering
    \includegraphics[width=0.98\columnwidth]{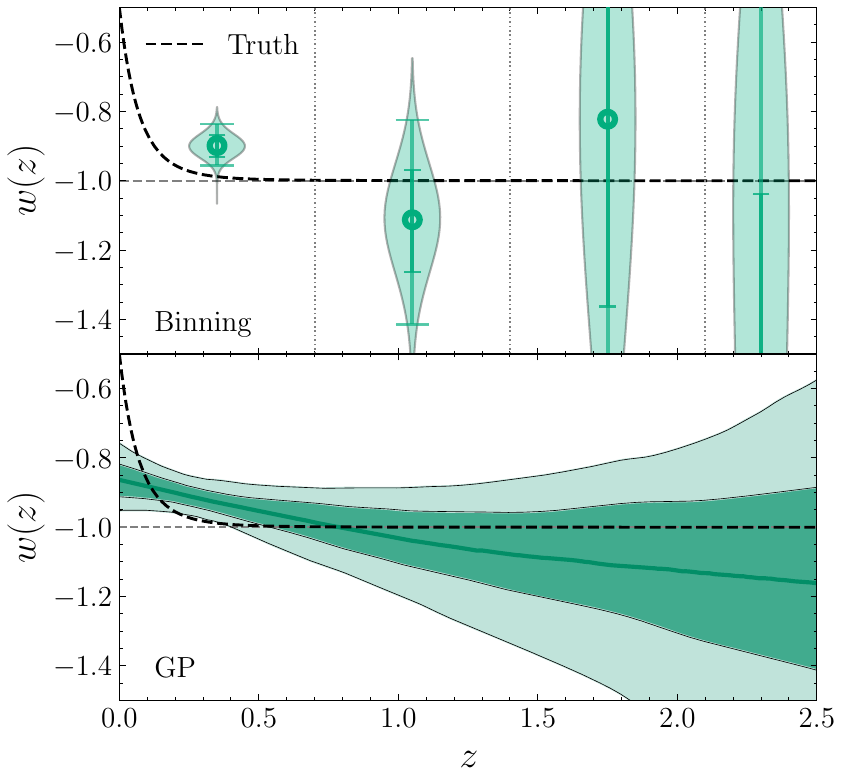}
    \caption{Reconstructions of $w(z)$ using binning and GP, for a selected extreme thawing case. In both reconstructions, the particular behavior used to generate the mock (shown in black dashed line) cannot be recovered well, given the implementation used.}
    \label{fig:xtreme_mock}
\end{figure}

In this appendix, we assess the robustness of the non-parametric methodologies used in the main text with a set of two mock datasets, each generated from a distinct dark energy model and comprising a combination of simulated DESI, PantheonPlus SNe~Ia, and CMB data. The first synthetic dataset assumes a \lcdm\ cosmology ($w_0=-1,w_a=0$), with the cosmological parameters set to their bestfit Planck values. The second one is a $w_0w_a\rm CDM$ realization, with $w_0=-0.75$ and $w_a=-0.85$. We refer to these as the $\Lambda$CDM mock and the $w_0w_a$CDM mock, respectively. While admittedly not an exhaustive sample of possible dark energy models, these different mocks span a range of physical behaviors that allow us to quantify the statistical uncertainties and possible biases introduced by the methods themselves.

\cref{fig:binning-fde-mocks} shows the binned $w(z)$ reconstruction in the case of 3 uniform bins between $z=0$ and $z=2.1$. The median values with $68\%$ and $95\%$ CL for the two mock datasets are shown. The true $w(z)$ for both mocks, plotted with a black dashed line,  lies mostly within the $1\sigma$ contours, sometimes straying into the $2\sigma$ range near the edges of each bin where, as expected, the three uniform bin scheme is not necessarily flexible enough to capture the precise behavior. The deviation from $\Lambda$CDM (shown in gray dashed line) is correctly detected at more than $2\sigma$ in the two lowest redshift bins in the case of the $w_0w_a$CDM Mock. \cref{fig:w_GP_mocks} shows that Gaussian process (GP) reconstructions of $w(z)$ for the two mock datasets. The figure includes the median values as well as $68\%$ and $95\%$ CL for both of the mock datasets. The true expected function lies well within the $1\sigma$ contours in each case.

\begin{figure}
    \centering
    \includegraphics[width=1\columnwidth]{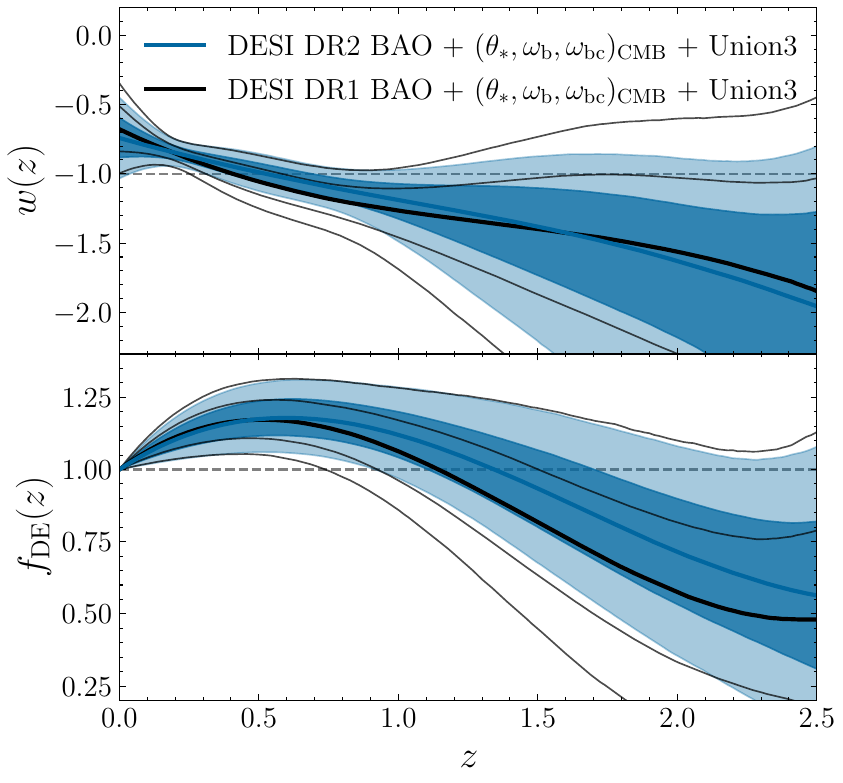}
    \caption{Comparison of the constraints obtained using DESI DR1 vs DR2 data, in combination with $(\theta_\ast,\omega_\mathrm{b},\omega_\mathrm{bc})_\mathrm{CMB}$ and Union3 measurements. The top panel shows a Chebyshev expansion of $w(z)$ as in \cref{eq:crossing_w}, while the bottom panel shows an expansion of $\fde(z)$ as in \cref{eq:crossing_fde} with $N=3$.}
    \label{fig:DR1-vs-DR2}
\end{figure}

In the interest of completeness, we also test an extreme case, falling in the thawing class of models, by preparing a third mock following~\cref{eq:thaw-algebraic} with $w_0=-0.5$ and $p=20$. As shown in~\cref{fig:xtreme_mock}, this demonstrates the possibility of limitations in the non-parametric implementation used here to pick up some extreme behaviors. We simply note that our implementation does have some limitations, and further improvements, such as investigating the impact of priors on hyperparameters and the choice of kernel, are left to future work.

Aside from the extreme case, the two approaches investigated here are seen to recover the simulated mock data well, without any significant bias detected in the mocks tested. Though not shown here, the same tests are performed for direct $\fde(z)$ reconstruction, with comparable analogous results. These non-parametric implementations are then applied without modification to the real data for the actual analysis.

\section{Comparison with DESI DR1}

\begin{figure}[b!]
    \centering
    \includegraphics[width=1\columnwidth]{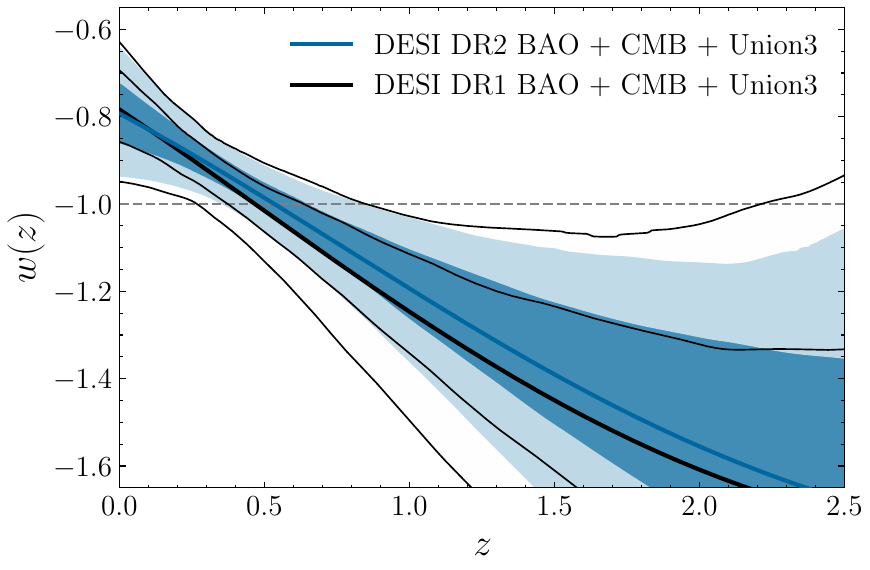}
    \caption{Comparison of GP reconstruction obtained using DESI DR1 vs DR2 data, in combination with CMB and Union3 measurements.}
    \label{fig:GP-DR1-vs-DR2}
\end{figure}

\Cref{fig:DR1-vs-DR2} compares the results of a Chebyshev expansion of \( w(z) \) and \( f_{\text{DE}}(z) \) using DR1 BAO vs DR2 BAO data, in combination with $(\theta_\ast,\omega_\mathrm{b},\omega_\mathrm{bc})_\mathrm{CMB}$ and Union3 measurements. \Cref{fig:GP-DR1-vs-DR2} presents a similar comparison for the GP reconstruction of  $w(z)$ using DESI in combination with CMB and Union3.  It is seen that while the main trends in the DE remain unchanged between the two data releases, the uncertainties in the reconstructions have significantly decreased with DR2. 

\end{document}